\documentclass[preprint,authoryear,12pt]{elsarticle}

\usepackage{amssymb}

\journal{Physics of the Dark Universe}

\begin{document}

\begin{frontmatter}



\title{Dark Energy Simulations}


\author{Marco Baldi}

\address{Excellence Cluster Universe\\
Boltzmannstrasse 2, 85748 - Garching bei M\"unchen, Germany}

\begin{abstract}
Cosmology is presently facing the deep mystery of the origin of the observed accelerated expansion of the Universe.
Be it a cosmological constant, a homogeneous scalar field,
or a more
complex inhomogeneous field possibly inducing effective modifications of the laws of gravity, such elusive physical entity
is indicated with the general term of ``Dark Energy".
The growing role played  by numerical N-body simulations in cosmological studies
as a fundamental connection between theoretical modeling and direct observations
has led to impressive advancements also in the development and application of specific algorithms
designed to probe a wide range of Dark Energy scenarios. Over the last decade, a large number
of independent and complementary investigations have been carried out in the field of
Dark Energy N-body simulations, starting from the simplest case of homogeneous Dark Energy models
up to the recent development of highly sophisticated iterative solvers for a variety of Modified Gravity theories.
In this Review {-- which is meant to be complementary to the general Review by \citeauthor{Kuhlen_Vogelsberger_Angulo_2012} published in this Volume --} I will discuss the range of scenarios for the cosmic acceleration that have been
successfully investigated by means of dedicated N-body simulations, and I will provide a broad summary of the
main results that have been obtained in this rather new research field. I will focus the discussion on a few selected
studies that have led to particularly significant advancements in the field, and I will provide a comprehensive 
list of references for a larger number of related works. Due to the vastness of the topic, the discussion will not 
enter into the finest details of the different implementations and will mainly focus on the outcomes of the various
simulations studies. Although quite recent, the field of Dark Energy simulations has witnessed
huge developments in the last few years, and presently stands as a reliable approach to the investigation of the 
fundamental nature of Dark Energy.
\end{abstract}


\end{frontmatter}


\section{Introduction}
\label{sec:intro}
After centuries of philosophical speculation about the origin and the physical properties of the Universe, at the beginning of the last century cosmology was finally allowed to become a proper scientific 
discipline with the development of Einstein's theory of General Relativity in 1915 \citep{Einstein_1915} and with the subsequent derivation of cosmological solutions to Einstein's field equations
by Friedmann in 1922 \citep{Friedmann_1922}. Less than a hundred years later, we are now provided with a well-established framework to study the properties of the Universe as a whole and to interpret
an ever increasing amount of high-quality observational data that allow to continuously improve the constraints on a few basic parameters that fully characterize our present standard cosmological model.

The latter is based on the assumption of homogeneity and isotropy of space encoded by the Copernican principle, and on the observation of the cosmic expansion that was first detected by Slipher and Hubble in the end
of the 1920's \citep{Slipher_1927,Hubble_1929}. This fundamental observation, which clearly indicated a time evolution of the Universe and posed the basis for the development of the Hot Big Bang cosmological
scenario, removed any motivation for the quest of static solutions to the field equations of General Relativity, and led Einstein to reject his own hypothesis \citep{Einstein_1917} of a cosmological constant term that could prevent
a static Universe from collapsing under its own self-gravity. 

The idea of a cosmological constant $\Lambda $ acting as a sort of ``repulsive force" and capable
to counteract the attractive pull of gravity was then disregarded for most of the century, until new observations 
of galaxy correlations at large scales \citep[][]{Maddox_etal_1990} started to indicate a tension with the predictions of a flat matter-dominated Universe. Finally, at the very end of the 20th century, the extraordinary discovery that the cosmic expansion is presently accelerating \citep{Riess_etal_1998,Perlmutter_etal_1999,Schmidt_etal_1998} suddenly revived the interest in the
cosmological constant as the simplest possible explanation for such new observational evidence.
Together with the wide range of astrophysical data supporting the existence of Cold Dark Matter (CDM) as the main fraction of the total cosmic mass \citep*[see e.g.][]{Bertone_Hooper_Silk_2005,Bergstrom_2012}, the discovery of the accelerated expansion represents
one of the observational pillars on which the presently accepted standard cosmological model is founded. \\

Despite the remarkable success of the simple original idea of a cosmological constant in describing the observed properties of the accelerating Universe -- as a consequence of which the standard model
takes the name of ``$\Lambda $CDM" cosmology -- the theoretical roots of such idea are yet poorly
defined and difficult to accommodate in the context of General Relativity and Quantum Field Theory \citep[see e.g.][]{Weinberg_1988}. As a matter of fact, the cosmological constant $\Lambda $ has to be highly fine-tuned with respect to the natural energy scales of the early Universe in order to provide the excellent fit to cosmological observations that presently still supports its success. For this reason, alternative
explanations for the observed cosmic acceleration have been proposed, and are generically
indicated with the term ``Dark Energy".

Dark Energy (DE) is then simply a label with which cosmologists indicate any physical mechanism
capable to provide an acceleration of the cosmic expansion compatible with our present
observational constraints. {Such} possible mechanisms -- which include the cosmological constant as the simplest option -- encompass a wide range of other alternative and more sophisticated possibilities. {These include, among others,} new
fields and interactions in the Universe, cosmological models with extra dimensions, modifications of General Relativity, local deviations from the Copernican principle, and backreaction effects of the formation of cosmic structures on the overall cosmic expansion \citep[for a general and recent review, see e.g.][]{Euclid_TWG}.\\

Most of the present efforts of theoretical and observational cosmologists are
devoted to the investigation of the DE phenomenon, with the aim to restrict the range of potentially
viable scenarios for the cosmic acceleration and to constrain their specific parameters. In such context, several ambitious
observational initiatives have been put in place worldwide to probe the nature of DE, and will provide complementary data
of unprecedented quality over the next decade. These include e.g.
the Dark Energy Survey \citep[DES,][]{DES}, the Hobby-Eberly Telescope Dark Energy EXperiment  
\citep[HETDEX,][]{HETDEX}, the Large Synoptic Survey Telescope \citep[LSST,][]{LSST} and the recently selected European Space Agency satellite mission Euclid \citep[][]{Euclid-r} that will be launched in 2020. Such large amount of data will have to be confronted with a wide variety of theoretical
proposals of ever increasing complexity and sophistication \citep[see e.g. the recent and comprehensive review of the Euclid collaboration,][]{Euclid_TWG} with the aim to detect possible specific observational
footprints identifying a particular DE candidate. As a matter of fact, the detection of any deviation from
the expected behavior of a cosmological constant would represent a breakthrough in our understanding
of the Universe and would open the way for the discovery of new physics.\\

The comparison between observational data and theoretical models of the Universe is however not a straightforward process. Besides the ever more complex procedures required to reduce raw data, quantify systematic errors, and extract
meaningful cosmological information from direct observations, one also needs to take into account the
corresponding difficulty of providing reliable theoretical predictions for the same observable quantities. {In fact, these
often require} to model highly nonlinear processes and involve the superposition of different physical mechanisms with potentially degenerate effects. 

In this respect, the use of numerical simulations to investigate the evolution of the Universe and
the formation of cosmic structures beyond the linear regime that is readily accessible to analytical computations has
proven to be an extremely valuable tool for the development of our understanding of the Cosmos.
This is already true for the simplest standard $\Lambda $CDM model, but it
becomes even more relevant for more complex DE scenarios for which one aims at identifying
small deviations from the standard predictions and looking for such small deviations in the data.
Significant progress has been made in the field of cosmological numerical simulations over the last decades, 
both due to the increase of the available computational power and to the development of efficient and sophisticated algorithms. These have allowed to study in detail the nature of Dark Matter and its role in driving
the growth of cosmic structures starting from the tiny density fluctuations generated in the early Universe
by the inflationary accelerated expansion, and to establish the CDM paradigm as the main
framework for the formation of galaxies and galaxy clusters {\citep[see e.g. the general Review by][included in the present Volume]{Kuhlen_Vogelsberger_Angulo_2012}}. More recently, and in particular after the
discovery of the cosmic acceleration, numerical simulations have also been used to test the nature of
DE, by employing ever more sophisticated implementations capable of capturing the characteristic features of several different and competing DE candidate models. Although this is a quite new and
rapidly developing field, numerical simulations of DE scenarios beyond the cosmological constant 
have now made sufficient progress
to deserve full consideration as a robust and reliable approach to the investigation of the DE phenomenon.
{Therefore, cosmological N-body simulations now stand} as an essential link between theoretical modeling and direct observations for any present and future collaborative initiative aimed at the study of the accelerated expansion of the Universe.\\

The present Review is meant to provide a broad overview on the developments and the results
achieved in the field of numerical simulations for different DE models. The focus will be more
concentrated on the conclusions reached by different simulation codes rather than on their numerical implementation details.
Also, due to the vastness of the topic, it will be clearly impossible to discuss most of the results presented in this work in full detail, and 
consequently this Review should be mainly taken as a general reference to address potentially interested readers to the relevant literature. {A more general Review on cosmological N-body simulations mostly focused on the study of Dark Matter properties has been recently compiled by \citeauthor*{Kuhlen_Vogelsberger_Angulo_2012} and can be found as a separate contribution to this Special Issue \citep[][]{Kuhlen_Vogelsberger_Angulo_2012}}.\\

The paper is organized as follows. In Section~\ref{sec:hist} I will present some historical
outline on the role played by cosmological N-body simulations in the investigation of the DE phenomenon; in Section~\ref{sec:models} I will briefly summarize the main classes of DE models
alternative to the standard cosmological constant; in Section~\ref{sec:background} I will review
recent results of N-body simulations for DE models that only modify the background expansion
history of the Universe with respect to $\Lambda $CDM; in Section~\ref{sec:inhomogeneous} I will then review the
results of simulations for models where the DE also directly alters the growth of cosmic structures due to its density perturbations or interactions. 
Finally, in Section~\ref{sec:concl}
I will provide a summary and drive my conclusions.

\section{Dark Energy and numerical simulations: some historical remarks}
\label{sec:hist}

Numerical N-body simulations have been very successfully employed over the last 
fifty years to study the properties and the formation processes of collapsed systems in the Universe, 
and significantly contributed to establish the Cold Dark Matter (CDM) paradigm as the standard scenario
for structure formation \citep[see e.g.][]{Aarseth_1963,Peebles_1970,White_1976b,Frenk_White_Davis_1983,Davis_etal_1985,White_etal_1987,Navarro_Frenk_White_1995,NFW,Klypin_etal_1999,Moore_etal_1999,Millennium,Aquarius,Angulo_etal_2012}.
However, cosmological simulations have also played a major role in the discovery and in the subsequent investigation
of the DE phenomenon. In fact, despite the undoubtable importance of the direct detection of the cosmic acceleration
by Perlmutter, Riess and Schmidt (recently recognized also by the award of the Physics Nobel Prize 2011) 
it is worth to remind that the first observational claim of a DE-dominated universe
came about ten years before from the comparison of the large-scale correlation of galaxies in the APM galaxy survey
with the predictions of N-body simulations \citep[][]{Maddox_etal_1990,Efstathiou_etal_1990}.

In particular, \citeauthor{Maddox_etal_1990} compared the correlation function extracted from the simulations of a CDM dominated Universe performed by \cite{White_etal_1987} with the APM
observational correlation function, and found a stark discrepancy between the two for large correlation angles, with the latter showing a higher
level of clustering at large scales as compared to the numerical predictions. Shortly after, \cite{Efstathiou_etal_1990} showed that such large discrepancy was removed when
comparing the data with simulations of a flat low-density Universe with $\Omega _{M}\approx 0.2$, where the missing energy for closure was given by a Cosmological Constant $\Lambda$.
Therefore, it seems not inappropriate to state that the first observational evidence of a DE-dominated Universe was actually derived from the outcomes of cosmological N-body simulations.

The connection between N-body simulations and DE investigations is then definitely not a new research field, 
although as a matter of fact it was only relatively recently that simulation codes suitable to explore a significant range of DE scenarios beyond the standard $\Lambda $CDM cosmological model started to be developed and applied. For long time, in fact, most of the efforts in numerical cosmology have been devoted to improve
the efficiency and the scalability of standard N-body algorithms for the $\Lambda $CDM scenario. {Such efforts have been mainly driven by} the aim to reach higher and higher levels of detail in the description of the properties of nonlinear structure formation, as well as to include in the integration scheme the effects of baryonic physics \citep[see e.g.][]{Teyssier_2001,Springel_Hernquist_2002,Duffy_etal_2010} and a wide range of astrophysical processes 
such as gas cooling, star formation, and feedback mechanisms from supernovae explosions and AGN activity \citep[see e.g.][]{Springel_Hernquist_2003a,Springel_Hernquist_2003b,Kay_etal_2002,Schaye_2004,Sijacki_etal_2007, DallaVecchia_Schaye_2008}. Alternatively, large N-body simulations of the standard $\Lambda $CDM scenario have also been used to develop and calibrate semi-analytic methods to populate simulated CDM halo catalogs
with realistic galaxy samples \citep[][]{White_Frenk_1991, Lacey_Cole_1993,Kauffmann_White_Guiderdoni_1993, Cole_etal_1994,Kauffmann_etal_1999, Somerville_Primack_1999,Springel_etal_2001,DeLucia_etal_2006}.

Both these approaches have driven spectacular progress in the understanding of galaxy formation and
evolution as well as in the capability of directly relating the outcomes of large numerical simulations to
real observations of galaxy and cluster populations. We are then now provided with a sophisticated and
robust numerical machinery for simulating the evolution of primordial density perturbations into a wide variety of possible observable quantities. Nonetheless, certainly due to the excellent fit that a simple
cosmological constant provides to most presently available data, all such developments have been pursued assuming a $\Lambda $CDM cosmology as the framework within which complex astrophysical
processes should take place. However, from a theoretical perspective 
the cosmological constant does not appear as a satisfactory explanation of the DE phenomenon, and 
a wide range of alternative scenarios have been proposed insofar, as already briefly mentioned above.
The attempt to include such alternative scenarios into the capabilities of N-body algorithms --
with the aim to investigate their effects on structure formation processes -- comes then as 
a natural further step in the connection between theoretical and observational cosmology.

\section{Dark Energy models}
\label{sec:models}

It is not surprising that the discovery of the accelerated expansion of the Universe triggered a 
great deal of theoretical attempts to provide a sensible explanation (possibly with a lower degree of fine-tuning
than the cosmological constant) to this mysterious phenomenon. New DE models are proposed almost on a daily basis (since June 1998, the number of papers containing the term``dark energy" in the title is about 3 thousand, corresponding to more than a paper every second day\footnote{data from www.arXiv.org}), and often do not differ
sufficiently from each other in their observational predictions to be possibly distinguished by presently available data.
A complementary approach to the development of specific DE scenarios based on different assumptions or on additional physical degrees of freedom with respect to the standard model, is that of parameterizing our ignorance
about the fundamental nature of DE with a few parameters quantifying possible deviations from the $\Lambda $CDM
behavior. In both cases, in order to obtain realistic predictions for observables that involve -- directly or indirectly -- the nonlinear evolution of cosmic structures, it is necessary to include the characteristic features of each specific model or parameterization into the algorithms of cosmological N-body solvers. \\

The range of available models and parameterizations is indeed quite
large, including violations of large-scale homogeneity and isotropy, new dynamical fields, effective or fundamental modifications of the laws of gravity, and extra dimensions. It clearly goes beyond the scope of the present Review
to present and discuss in detail the main features of all these different extensions of the standard model, for which
I refer to some specific recent reviews \citep[see e.g.][]{Copeland_Sami_Tsujikawa_2006,Tsujikawa_2010,Sapone_2010, Kunz_2012,Euclid_TWG}. For what concerns the topics discussed in this work, a sensible classification of
DE scenarios should be based on the way in which different models can possibly affect the processes of structure formation, and in particular on how they are expected to modify the nonlinear collapse of gravitationally bound systems. Following this
general principle, we can define three main categories of DE models: {\bf Homogeneous DE fields}, {\bf Inhomogeneous DE fields}, and {\bf Large-Void inhomogeneous cosmologies}. Far from trying to be complete, I will briefly summarize the main features
and the most common examples of these three different classes in the remainder of this section.

\subsection{Dark Energy as a homogeneous field}

For a homogeneous and isotropic Universe described
by a Friedmann-Lema\^itre-Robertson-Walker (FLRW) metric:
\begin{equation}
ds^2 = -c^{2}dt^{2} + a(t)\left\{ \delta _{ij}dx^{i}dx^{j}\right\}
\end{equation}
where the time dependence of the line element is confined in the scale factor $a(t)$,
the background evolution of the Universe is encoded by the Hubble function $H(a)\equiv \dot{a}/a$ which describes
how the expansion rate changes as a function of time. Here an overdot represents a derivative with respect to the cosmic time
$t$ and I assume the scale factor $a$ to be normalized at unity today.
The Hubble function is then related to the relative abundance of the
different constituents of the Universe through the Friedmann equation:
\begin{equation}
\label{Friedmann}
\frac{H^{2}(a)}{H_{0}^{2}} = \Omega _{\rm M}a^{-3} + \Omega _{\rm r}a^{-4} + \Omega _{\rm K}a^{-2} + \Omega _{\rm DE}\exp \left\{ -3 \int_{1}^{a}\frac{1+w(a')}{a'}da'\right\}\,,
\end{equation}
where $\Omega _{\rm i}$ is the energy density of the i-th component of the Universe at the present time in units of the critical density $\rho _{\rm crit}\equiv 3H_{0}^{2}/8\pi G$, and the different components considered are matter (M), radiation (r), curvature (K) and Dark Energy (DE).
The equation of state parameter $w(a)$ quantifies the ratio between pressure and energy densities of the DE component, and is allowed to be time-dependent.
As one can see from Eq.~\ref{Friedmann}, a cosmological constant corresponds to a constant value of $w=-1$, which implies a constant energy density of DE throughout the whole expansion history of the Universe. 
On the other hand, different constant or time-dependent values of the equation of state parameter would imply some
evolution of the DE density and would consequently affect the expansion rate $H(a)$.\\

In the late Universe (i.e. sufficiently after matter-radiation equivalence at $z_{\rm eq}\approx 3\times 10^{3}$) the growth of linear density perturbations at sub-horizon scales is described, in the Newtonian gauge and in Fourier space, by the following evolution equation:
\begin{equation}
\label{perturbations1}
\ddot{\delta }_{\rm M} + 2H\dot{\delta }_{\rm M}= 4\pi G \left(\rho _{\rm M}\delta _{\rm M} + \rho _{\rm DE}\delta _{\rm DE}\right)\,,
\end{equation}
where $\delta _{\rm M,DE}\equiv \delta \rho _{\rm M,DE} /\rho _{\rm M,DE}$ is the density contrast of the matter and DE components. If one assumes that the DE field does not appreciably cluster at sub-horizon scales, i.e. if the DE component is homogeneous over the whole causally connected Universe, the second term on the right-hand side of Eq.~\ref{perturbations1} vanishes at all times since $\delta _{\rm DE}=0$, and the only impact that DE can have on structure formation processes comes through the Hubble friction term $2H\dot{\delta }_{\rm M}$ appearing on the right-hand side of Eq.~\ref{perturbations1}. Therefore, a non-standard yet homogeneous DE component characterized
by an equation of state parameter $w\neq -1$ will affect the evolution of density perturbations only in an indirect way through a different expansion history. Nevertheless, the impact of this class of scenarios on the linear and nonlinear
evolution of structures can still be substantial as the gravitational collapse of density perturbations will occur at different epochs depending on the evolution of the linear growth factor. \\

The homogeneity of the DE field at sub-horizon scales can either be taken as an assumption for a wide range of phenomenological
parameterizations of the DE background evolution, or can arise as an intrinsic feature of DE scenarios based on
the dynamical evolution of a light scalar field $\phi $ as in the case of {\em Quintessence} \citep{Wetterich_1988,Ratra_Peebles_1988}, {\em k-essence} \citep{kessence}, {\em Phantom} \citep[][]{Caldwell_2002} and {\em Quintom} \citep{Feng_Wang_Zhang_2005} DE models. The latter are generally characterized by a scalar field sound speed $c_{s}^{2}$ equal or comparable to the speed of light $c$,
thereby suppressing perturbations of the DE density within the cosmic horizon, {while DE perturbations remain frozen to a constant amplitude at super-horizon scales. This implies that 
density fluctuations in the DE field are in any case present at scales comparable to the cosmological horizon even for scalar field models with a high sound speed $c_{s}^{2}\approx c$ \citep[see e.g.][]{Ma_etal_1999,Bean_Dore_2004,Weller_Lewis_2003,Bartolo_etal_2004a}. In particular, DE perturbations might still change the large-scale shape of the matter power spectrum, thereby affecting the initial conditions for structure formation \citep[see e.g.][]{Ma_etal_1999,Alimi_etal_2010}. Nevertheless, the impact of horizon-scale DE perturbations on the nonlinear evolution of structures for this class of models is rather small and can be expected to play a significant role only for very large cosmological simulations with a comoving size comparable to the cosmic horizon. For this reason, assuming the homogeneity of the DE field for this class of scenarios represents a valid approximation for a wide range of numerical setups, while only the recent development of extremely large N-body simulations for DE cosmologies \citep[see e.g.][]{Alimi_etal_2010,Rasera_etal_2010,Alimi_etal_2012} has required to carefully take into account the presence of DE perturbations in the initial conditions.}\\

As a general example {of scalar field DE models}, the dynamic equation
of a {\em Quintessence} scalar field is described by a homogeneous Klein-Gordon equation
\begin{equation}
\ddot{\phi } + 3H \dot{\phi } + \frac{dV}{d\phi } = 0
\end{equation}
where $V(\phi )$ is a self-interaction potential and where the DE density is given by $\rho _{\rm DE} = \dot{\phi }^{2}/2 + V(\phi )$. Different choices of the function $V(\phi )$ will then determine different evolutions of the DE density and will affect in specific ways the expansion history and consequently the growth of cosmic structures. Some of the most widely used forms
of the function $V(\phi )$ include {\em runaway} potentials as e.g. an inverse power-law \citep{Ratra_Peebles_1988}
\begin{equation}
\label{RP}
V(\phi ) = A\phi ^{-\alpha }\,,
\end{equation}
or an exponential \citep{Lucchin_Matarrese_1984,Wetterich_1988}:
\begin{equation}
\label{EXP}
V(\phi ) = Ae^{-\alpha \phi }\,,
\end{equation}
as well as {\em confining} functions as e.g. the SUGRA potential \citep{Brax_Martin_1999} arising naturally within supersymmetric theories of gravity:
\begin{equation}
\label{SUGRA}
V(\phi ) = A\phi ^{-\alpha }e^{\phi ^{2}/2}\,.
\end{equation}
In Eqs.~\ref{RP}-\ref{SUGRA} the scalar field $\phi $ has been expressed in units of the reduced Planck mass $M_{\rm Pl}=1/\sqrt{8\pi G}$ and is therefore dimensionless. These three potentials represent the most widely used choices
for {\em Quintessence} models as they provide viable expansion histories and scaling solutions that
make the cosmological evolution largely independent from the scalar field initial conditions \citep[see e.g.][]{Ferreira_Joyce_1998},
and have been widely investigated through N-body simulations. 

Cosmological models characterized by a vector field, rather than a scalar, playing the role of DE have also been recently proposed \citep[][]{BeltranJimenez_Maroto_2008}.
In such scenarios, cosmic acceleration is driven by the kinetic energy of the vector field, without resorting on any arbitrary choice of a potential function. Despite the vector nature of the DE field, the energy density of its spatial components dilutes faster than matter with the cosmic expansion, and is therefore negligible for the evolution of the late Universe. These models therefore behave similarly to scalar field DE cosmologies, inducing a modified background expansion history without significant sub-horizon perturbations of the DE density, although the fundamental mechanism behind the accelerated expansion is different from standard Quintessence scenarios. \\

A different possibility, already mentioned above, is to assume {\em a priori} the homogeneity of the DE field and describe its
time evolution by phenomenological parameterizations of the DE equation of state parameter $w(a)$.
Several different options have ben proposed in the last years, either based on the behavior of $w(a)$ at low redshifts, as
for the case of the so-called Chevallier-Polarski-Linder  \citep[CPL,][]{Chevallier_Polarski_2001,Linder_2003} parameterization 
\begin{equation}
\label{CPL}
w(a) = w_{0} + w_{a}(1-a)
\end{equation}
where $w_{0}$ and $w_{a}$ are constants, or assuming as main parameters the relative abundance of DE at the present epoch ($1-\Omega _{\rm M}$) and at early times ($\Omega _{\rm EDE}$), as for the case of the Early Dark Energy parameterization of \cite{EDE}:
\begin{equation}
\label{EDE}
w(a)=\frac{w_{0}}{1+b\ln (1/a)}\,, \quad {\rm where} \quad b = \frac{3w_{0}}{\ln \frac{1-\Omega _{\rm EDE}}{\Omega _{\rm EDE}} + \ln \frac{1-\Omega _{\rm M}}{\Omega _{\rm M}}}\,.
\end{equation}

All these different scenarios and parameterizations significantly affect the growth of density perturbations both in the linear and nonlinear regimes, and have been extensively investigated with N-body simulations over the last decade, as will be discussed in Section \ref{sec:background}.

\subsection{Dark Energy as an inhomogeneous field}

If DE is associated to some new physical degree of freedom rather than to a cosmological constant, it is natural
to consider also its spatial fluctuations and its possible interactions with other components of the Universe. The assumption
of homogeneity discussed above might therefore be a reasonable approximation {at sufficiently small scales} 
for a wide range of DE scenarios characterized
by a large sound speed of the DE fluid and by the absence of substantial direct interactions of DE besides gravity, but certainly
fails in describing the most general possible case of a DE field beyond $\Lambda $. A large variety of DE models featuring
significant perturbations at sub-horizon scales and/or substantial interactions with matter or gravity have been proposed in the last years, and generically form the class of inhomogeneous DE cosmologies.

A first example of such models is given by the {\em Clustering DE} scenario \citep[e.g.][]{Creminelli_etal_2009,Creminelli_etal_2010,Sefusatti_Vernizzi_2011}, where a general k-essence scalar field 
is simultaneously characterized by an equation of state parameter generally different from $-1$ and by a ``cold" sound speed $c_{s}^{2}\approx 0$. As a consequence, DE can cluster also below the horizon and source gravitational potentials at scales relevant 
for the formation of cosmic structures. This corresponds to the case $\delta _{\rm DE}\neq 0$ that was discarded above under the assumption of homogeneity, which implies that the net potential for the growth of CDM density perturbations will include
also the contribution of DE perturbations, according to the full form of Eq.~\ref{perturbations1}, for which one can write:
\begin{equation}
\label{perturbations2}
\ddot{\delta }_{\rm M} + 2H\dot{\delta }_{\rm M}= 4\pi G \rho _{\rm M}\left( \delta _{\rm M} + \frac{\Omega _{\rm DE}}{\Omega _{\rm M}}\delta _{\rm DE}\right) \,.
\end{equation}
From Eq.~\ref{perturbations2} one can clearly see that  DE perturbations will substantially affect the evolution of CDM structures only at late times, when the background DE density becomes important as compared to the CDM one. Also, it is interesting to notice how an observer
ignoring the clustering properties of DE could interpret the evolution of perturbations determined by Eq.~\ref{perturbations2} as 
a modification of gravity emerging in the late Universe. 
The example of Clustering DE models then already clearly shows how a fundamental distinction between a DE degree of freedom and
a modification of the laws of gravity at astrophysical scales results impossible whenever one allows for spatial perturbations in the DE field: the specific
clustering properties of a DE field can in general mimic deviations from the expected behavior of standard gravitational instability processes induced by a modified force law. \\

The fundamental degeneracy between these two different perspectives becomes even more evident for the case of DE fields
featuring direct interactions with matter, for which a formal correspondence to modified theories of gravity through a conformal
transformation of the metric can be explicitly demonstrated \citep[see e.g.][]{Pettorino_Baccigalupi_2008}. {\em Interacting DE} models and {\em Modified Gravity} theories therefore represent a unique class of cosmological scenarios beyond $\Lambda $CDM for which structure formation processes are in principle modified both by a non-standard evolution of the background expansion history and by the specific clustering and interaction properties of the new degrees of freedom associated to the DE sector of the Universe. Such cosmologies are in fact generically characterized by the existence of fifth-forces mediated by these new degrees of freedom, whose spatial range and universality depend on the specific model under consideration. A detailed overview and classification of Interacting DE and Modified Gravity models goes beyond the scope of the present Review, and I refer the interested reader to some excellent recent publications which provide a self-consistent and comprehensive overview on these scenarios \citep[see e.g.][and references therein]{Tsujikawa_2010,DeFelice_Tsujikawa_2010,Euclid_TWG}. For what concerns the aims of this work, it is sufficient to identify
the few main features that determine how different specific models belonging to this class of cosmologies directly affect the growth of density perturbations in the linear and nonlinear regimes.\\

As already mentioned above, a general feature of Interacting DE and Modified Gravity models is the existence of
a fifth-force of nature, mediated by the scalar degree of freedom associated to DE. In the most general case, such fifth-force can be described as an additional
term in the acceleration equation of a massive test particle representing a fluid element of a given cosmic component $i$:
\begin{equation}
\label{fifth-force}
\dot{\vec{v}}_{i} = -\vec{\nabla }\Phi - \beta _{i}(\phi )\vec{\nabla }\delta \phi
\end{equation}
where the standard gravitational potential $\Phi $ is determined by the Poisson equation:
\begin{equation}
\label{poisson}
\nabla ^{2}\Phi = 4\pi G \sum_{j}\rho _{j}\delta _{j}
\end{equation}
with $j$ ranging over all the different clustering components of the Universe. The additional scalar potential $\delta \phi $ obeys a modified non-linear Poisson equation of the form:
\begin{equation}
\label{modified_poisson}
\nabla ^{2}\delta \phi = F(\delta \phi ) + \sum_{j}8\pi G \beta _{j}(\phi ) \delta _{j}\,,
\end{equation}
with $F$ a generic function of the scalar field perturbation $\delta \phi $. As one can see from Eqs.~\ref{fifth-force}-\ref{modified_poisson}, the choice of the coupling functions $\beta _{i}(\phi )$ and the form of the function $F(\delta \phi )$ will
determine the configuration of the scalar perturbations $\delta \phi $ and the related fifth-force experienced by massive particles. The formulation presented above and described by Eqs. \ref{fifth-force}-\ref{modified_poisson} is rather general,
and covers a wide range of different models of Interacting DE and Modified Gravity. \\

As a first main classification of such scenarios, one can then start distinguishing between models featuring a {\em universal} coupling (i.e. $\beta _{i}(\phi ) = \beta (\phi )\, \forall i$) and models with {\em species-dependent} couplings. The former case, corresponding to modified gravity theories as e.g. $f(R)$ gravity \citep[see e.g.][and references therein]{Hu_Sawicki_2007,DeFelice_Tsujikawa_2010}, {\em Extended Quintessence} models \citep[][]{Baccigalupi_Matarrese_Perrotta_2000,Perrotta_etal_2000,Pettorino_etal_2005,Pettorino_Baccigalupi_2008}, higher-dimensional theories of gravity as e.g. {\em DGP} \citep{Dvali_Gabadadze_Porrati_2000}, or the recently proposed {\em Galileon} \citep{Nicolis_Rattazzi_Trincherini_2009}, {\em Symmetron} \citep[][]{Hinterbichler_Khoury_2010,Hinterbichler_etal_2011} and {\em Dilaton} \citep[][]{Gasperini_Piazza_Veneziano_2002} cosmologies, requires that
the fifth-force be suppressed in high-density environments in order to evade solar system constraints on possible deviations
from General Relativity \citep[see e.g.][]{Bertotti_Iess_Tortora_2003,Will_2005}. This suppression can be realized with
a variety of screening mechanisms, as e.g. the {\em Chameleon} \citep[][]{Khoury_Weltman_2004}, the {\em Vainshtein} \citep[][]{Vainshtein_1972,Deffayet_etal_2002} or the {\em Symmetron} \citep[][]{Hinterbichler_Khoury_2010}, which all rely on relatively large  fluctuations ($|\delta \phi |\sim \mathcal{O}(1)$ or larger) of the scalar field (or of its derivatives) between high-density regions and the cosmic low-density environment. 
Such large perturbations can arise e.g. when nonlinearities are present in the function $F(\delta \phi )$ appearing in Eq.~\ref{modified_poisson}, which then requires quite
sophisticated algorithms to be properly solved for an arbitrary matter distribution $\delta _{\rm M}(t\,,\vec{x})$ within newtonian N-body codes, as will be discussed in Section~\ref{sec:inhomogeneous}. This is for instance the case of $f(R)$ theories of gravity in the \citeauthor{Hu_Sawicki_2007} parameterization, for which $\delta \phi = f_{R}\equiv df(R)/dR$ and $F(\delta \phi ) = F(f_{R}) \propto R(f_{R}) - \bar{R}$ with the relation between $f_{R}$ and $R$ given by:
\begin{equation}
f_{R} = -n\frac{c_{1}}{c_{2}^{2}}\left(\frac{m^{2}}{R}\right) ^{n+1}\,,
\end{equation}
with $n$, $c_{1}$ and $c_{2}$ constants.

On the other hand, if one allows for {\em non-universal} couplings \citep[as first proposed by][]{Damour_Gibbons_Gundlach_1990}, solar system constraints can be easily evaded without resorting on any screening mechanism by simply assuming the coupling to baryons $\beta _{b}(\phi )$ to be highly suppressed. This second option corresponds to the general class of {\em Coupled DE} models where a non-vanishing coupling to CDM particles \citep[][]{Wetterich_1995,Amendola_2000,Amendola_2004} or to massive neutrinos \citep[as for the {\em Growing Neutrino} scenario, ][]{Amendola_Baldi_Wetterich_2008} provides viable cosmological expansion histories and a possible solution to the fine-tuning problems of the cosmological constant. For this class of models, the function $F(\delta \phi )$ in Eq.~\ref{modified_poisson} is related to the derivative of the scalar self-interaction potential $dV/d\phi $, and for sufficiently flat potentials (which are anyway required in order to provide an accelerated expansion of the Universe) can be safely discarded compared to the 
matter density perturbations, such that Eq.~\ref{modified_poisson} reduces to:
\begin{equation}
\nabla ^{2}\delta \phi \approx \sum_{j}8\pi G \beta _{j}(\phi ) \delta _{j}
\end{equation}
and for the case where only one species is coupled to DE one gets:
\begin{equation}
\nabla ^{2}\delta \phi \approx 8\pi G \beta _{i}(\phi ) \delta _{i} = 2\beta _{i}(\phi )\nabla ^{2}\Phi _{i}
\end{equation}
where $\Phi _{i}$ is the standard gravitational potential generated by the coupled matter component $i$.
From the previous equation, one immediately gets that $\delta \phi \approx 2\beta _{i}(\phi )\Phi _{i}$ and therefore 
from Eq.~\ref{fifth-force} the fifth-force acting on a coupled particle will be simply proportional to standard gravity
by a factor $2\beta _{i}^{2}$. More general scenarios featuring a coupling with two \citep{Baldi_2012a} or multiple \citep{Brookfield_VanDeBruck_Hall_2008} CDM fluids have also been recently proposed, for which the previous arguments apply separately to the fifth-force generated by each individual coupled component. 

For the case of a non-universal interaction between DE and other fluids in the Universe, an additional acceleration term appears in Eq.~\ref{fifth-force} as a consequence of momentum conservation in the coordinate frame of the minimally coupled species \citep[i.e. those species for which the coupling to DE vanishes, see e.g.][]{Amendola_2000,Maccio_etal_2004,Baldi_etal_2010}. Such additional term is in general proportional to the velocity vector of a test particle and has been therefore termed ``friction" or ``drag" term in the literature. The full acceleration equation of a coupled particle in the Einstein frame for Coupled DE models with a non-universal coupling then reads:
\begin{equation}
\label{fifth-force-fric}
\dot{\vec{v}}_{i} = \beta _{i}(\phi )\dot{\phi }\vec{v}_{i} - \vec{\nabla }\Phi - 2\beta _{i}(\phi )\sum_{j}\beta _{j}(\phi )\vec{\nabla }\Phi _{j}
\end{equation}
which for a self-consistent N-body implementation requires to separately solve for the gravitational potential of each differently-coupled matter component of the Universe. It is also interesting to notice here how the sign of the friction term depends on the relative signs of the scalar field background velocity $\dot{\phi }$ and of the coupling function $\beta _{i}(\phi )$. This peculiar form of the friction term can determine a quite broad phenomenology of interacting DE
models at the level of linear and nonlinear structure formation, as will be discussed in Section~\ref{sec:inhomogeneous} below.

\subsection{Large-Void models}

A further possibility to account for the accelerated expansion of the Universe without invoking a Cosmological Constant (and in this specific case
even without resorting on any other DE field) is to drop the assumption of large-scale homogeneity encoded in the Copernican Principle
and consider the possibility that the observed accelerated expansion be just an apparent effect due to a strong local deviation from homogeneity \citep[see e.g.][]{Mustapha_Hellaby_Ellis_1997, Tomita_2001,Wiltshire_2007,GarciaBellido_Haugboelle_2008}.
In particular, an observer sitting near the center of a very large underdensity would observe an apparent acceleration of the Universe due to
the different expansion rate of the void at different distances from its geometrical center. This class of scenarios goes under the name of Large-Void
or Lema\^itre-Tolman-Bondi (LTB) cosmologies as they derive from the general spherically symmetric space-time metric first studied by Lema\^itre, Tolman and Bondi \citep[][]{Lemaitre_1933,Tolman_1934,Bondi_1947}:
\begin{equation}
ds^{2} = -dt^{2}+\frac{A'^{2}(r,t)dr^{2}}{1-k(r)}+A^{2}(r,t)\left( d\theta ^{2} + \sin ^{2}\theta d\phi ^{2}\right) \,,
\end{equation}
where $A$ and $A'$ are functions of time and of the radial coordinate from the center of symmetry of the system.

Although LTB models require a very large size  of the density void ($\sim 2$ Gpc or larger) in order to possibly explain the observed
accelerated expansion of the Universe without resorting to any additional DE field, they have attracted significant interest in the last years 
due to their simplicity and to the wide range of possible observational features that they provide and that could become directly observable 
with the next generation of surveys \citep[see e.g.][]{Quercellini_etal_2010}.

Viable large-void LTB cosmologies can be described by a four-parameters model of the void density profile $\Omega _{\rm M}(r)$ and of the
radial Hubble rate $H_{0}(r)$ according to the equations \citep[see][for more details]{GarciaBellido_Haugboelle_2008}:
\begin{eqnarray}
\Omega _{\rm M}(r) = 1 + \left( \Omega _{\rm in} - 1\right) \left( \frac{1 - \tanh \left[ \left( r - r_{0}\right) /2r \right] }{1 + \tanh \left[ r_{0}/2\Delta r\right] } \right) \,, \\
H_{0}(r) = H_{0} \left[ \frac{1}{\Omega _{\rm K}(r)} - \frac{\Omega _{\rm M}(r)}{\sqrt{\Omega _{\rm K}^{3}(r)}}\sinh ^{-1}\sqrt{\frac{\Omega _{\rm K}(r)}{\Omega _{\rm M}(r)}} \right] \,, 
\end{eqnarray}
where the four free parameters are the overall expansion rate $H_{0}$, the underdensity at the center of the void $\Omega _{\rm in}$, the radius of the void $r_{0}$, and the transition width of the void profile $\Delta r$ which defines how the profile matches from the inner value $\Omega _{\rm in}$ and the density
parameter at infinity which is assumed to be $\Omega _{\rm M}(\infty )=1$. Such class of models affects the growth of density perturbations due to the 
space-dependence of the cosmic density $\Omega _{\rm M}(r)$ which for very large voids will still be approximately constant over the scales of density perturbations collapsing into bound structures before the present epoch, but will significantly vary over different regions of the presently observable Universe. Despite large-void LTB models have recently started to show some tension with geometric probes of the expansion history of the Universe \citep[][]{Zumalacarregui_etal_2012}, the study of their effects on structure formation processes with the aim to identify possible observational footprints of a large void in the statistical properties of large-scale structures has recently attracted significant interest, and will be briefly discussed in Section~\ref{sec:LTB}.

\section{Simulating a Dark Energy background expansion}
\label{sec:background}

I now move to discuss how the different cosmological scenarios beyond $\Lambda $CDM that were introduced in Section~\ref{sec:models} have been investigated by means of dedicated N-body simulations for what concerns their effects on the formation of nonlinear cosmic structures.
I start such review from the simplest case of homogeneous DE models for which, as I explained above, the only effect on the growth of
density perturbations comes from a modified background expansion that changes the linear growth factor through the Hubble friction term of
Eq.~\ref{perturbations1}, {unless the simulated volume is so large to require a proper sampling of DE perturbations at scales comparable to the cosmic horizon. Consequently, cosmological simulations aiming at studying the evolution of structures in the context of these scenarios need to implement in their numerical algorithms only a proper modification of the expansion history $H(z)$.} \\

{The first simulations of homogeneous DE models with a constant equation of state parameter $w\neq -1$ have been performed by \cite{Ma_etal_1999} using a Particle-Particle/Particle-Mesh code to evolve $128^{3}$ particles within a periodic cosmological box of $100$ Mpc aside. The work of \citeauthor{Ma_etal_1999} focuses mainly on the detailed shape of the nonlinear matter power spectrum in constant-$w$ DE models, providing a fitting formula based on the specific growth factor of the different DE cosmologies. Soon after, \cite{Bode_etal_2001} performed a large suite of N-body simulations with $512^{3}$ particles in a $1$ Gpc periodic box for a variety of cosmologies, including also one DE model with $w=-2/3$, and investigated the evolution of the cluster mass function in the different scenarios, finding that the DE model shows a slower evolution of the cluster abundance, thereby determining a larger number of clusters at high redshift when a common normalization of the linear perturbations amplitude at $z=0$ is assumed.}

\begin{figure*}
\includegraphics[width=2.7in]{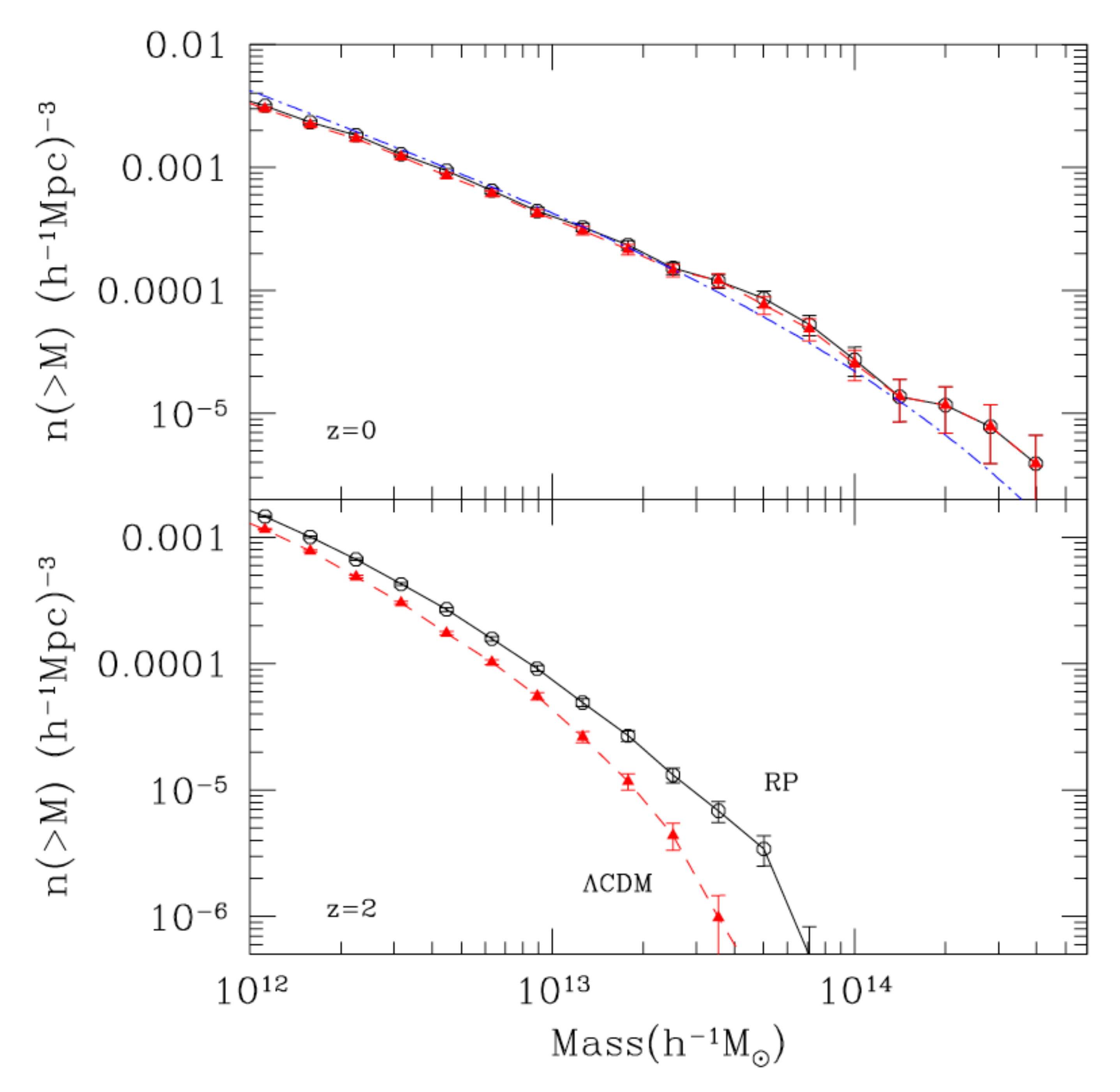}
\includegraphics[width=2.7in]{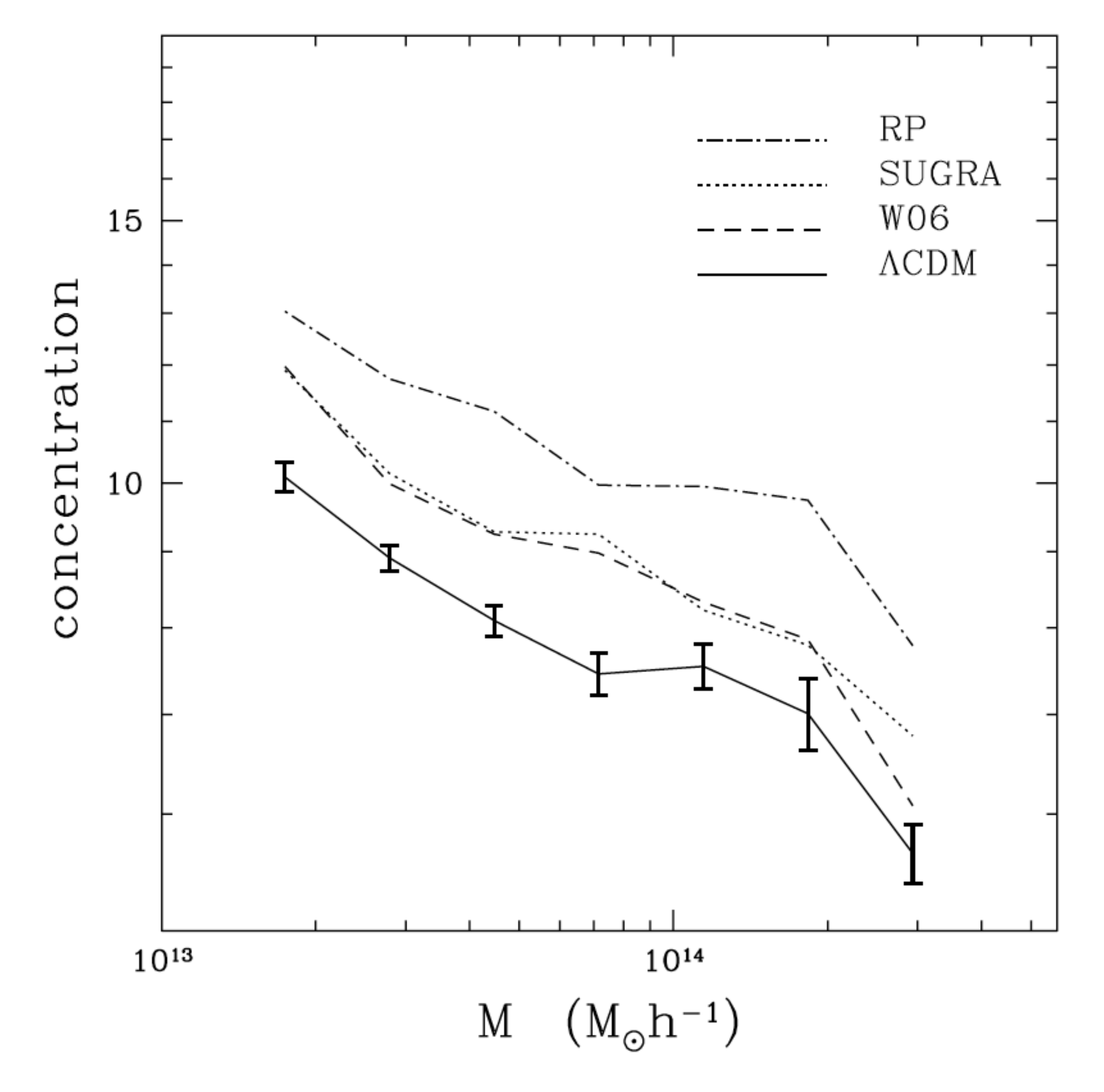}
\caption{{\em Plots from \cite{Klypin_etal_2003}} - {\em Left panel:} Comparison of the HMF for a $\Lambda $CDM cosmology (red) and a Quintessence model with an inverse power-law self-interaction potential (black) as extracted from N-body simulations with a box size of $160$ Mpc$/h$ and a mass resolution of $m=2\times 10 ^{10}$ M$_{\odot}/h$. The other DE scenarios considered by \citeauthor{Klypin_etal_2003} lie in between these two extreme models. The mass functions are practically indistinguishable at $z=0$ but at higher $z$ the DE cosmologies show a higher number of massive halos as compared to $\Lambda $CDM. {\em Right panel:} The concentration-mass relation at $z=0$ computed from the same set of simulations considered in the {\em left panel}. CDM halos in DE cosmologies result more concentrated due to their earlier formation epoch.}
\label{fig1}
\end{figure*}

{The first simulations of homogeneous DE models with a variable equation of state parameter $w\neq {\rm const.}$ have been performed
by \cite{Klypin_etal_2003} using a modified version of the Adaptive Mesh Refinement (AMR) code {\small ART} \citep[developed by][]{Kravtsov_Klypin_Khokhlov_1997}. In their work, \citeauthor{Klypin_etal_2003} investigated a few test models with either a constant equation of state $w > -1$ or a variable equation of state $w(a)$}
corresponding to the dynamical evolution of a Quintessence field with the inverse power-law and SUGRA potentials of Eqs.~\ref{RP} and \ref{SUGRA}. For their simulations, \citeauthor{Klypin_etal_2003} adopted a common normalization of the linear power spectrum of all the different cosmologies with the standard $\Lambda $CDM value of $\sigma _{8}$ at $z=0$, {similarly to what previously done by \citeauthor{Bode_etal_2001}} As we will see later on, the choice of the linear normalization
is a critical issue in the comparison of different cosmological scenarios with N-body simulations. The outcomes of these first runs
showed that no significant difference was present at $z=0$ among the various models in several observable quantities like the nonlinear matter power spectrum $P(k)$, the CDM halo mass function $N(>M)$, and the circular velocity function (i.e. the number of halos as a function of their maximum circular velocity). 
A significant scatter among the models could instead be noticed at higher redshifts, with the non-standard DE cosmologies systematically showing a higher number of halos as compared to $\Lambda $CDM both in the halo mass function and in the circular velocity function, with an enhancement increasing with halo mass (see Fig.~\ref{fig1}, left panel). Furthermore, the DE models did also show a higher amplitude of the matter power spectrum at all scales for $z>0$, consistently with the slower growth rate induced by the background scaling of the DE density. 

In the same work, \citeauthor{Klypin_etal_2003} also investigated the
inner structure of CDM halos in their various DE cosmologies by means of high-resolution zoom re-simulations of some of the most massive halos identified in the basic cosmological runs. This allowed them to show that the density profiles of CDM halos in homogeneous DE models still follow a Navarro-Frenk-White \citep[NFW,][]{NFW} profile
\begin{equation}
\frac{\rho (r)}{\rho _{crit}} = \frac{\delta ^{\star }}{(r/r_{s})\cdot (1+r/r_{s})^{2}}\,,
\end{equation}
but are systematically more concentrated than their $\Lambda $CDM counterparts, with a smaller value of the scale radius $r_{s}$. According to the concusions of \citeauthor{Klypin_etal_2003} this effect is likely due to the earlier formation redshift of halos in the DE scenarios, a picture which is again
consistent with a slower growth rate of density perturbations for models with a common linear perturbations normalization at the present epoch. 
Such effect determines a higher normalization for the concentration-mass relation in DE cosmologies as compared to $\Lambda $CDM (see Fig.~\ref{fig1}, right panel).
Additionally, they also found that the DE cosmologies systematically show a higher number of CDM satellite halos within massive collapsed structures, and that this effect seems to correlate with the higher circular velocity of CDM main halos in DE models as compared to $\Lambda $CDM.\\

A complementary approach was followed soon after by \cite{Linder_Jenkins_2003}, who investigated DE models with a parametrized equation of state $w(a)$ using the CPL parameterization of Eq.~\ref{CPL} with $w_{0}$ and $w_{a}$ chosen to best fit
Quintessence models with a SUGRA potential. With this approach, \citeauthor{Linder_Jenkins_2003} ran a series of cosmological simulations with a modified version of the Tree-PM code {\small GADGET} \citep[][]{gadget} within a somewhat larger box size as compared to \cite{Klypin_etal_2003} in order to better sample the high-mass tail of the halo mass function. This study also adopted a common normalization of the different models to the same $\sigma _{8}$ at $z=0$, and found consistent results with the {earlier} outcomes of \citeauthor{Klypin_etal_2003}: no significant deviations from the standard $\Lambda $CDM model at $z=0$, and a systematically larger abundance of CDM halos -- especially at large masses -- for the DE cosmologies at higher redshifts as compared to $\Lambda $CDM.
Additionally, \citeauthor{Linder_Jenkins_2003} showed that the the standard fitting formulae for the Halo Mass Function (HMF), and in particular the \cite{Jenkins_etal_2001} formula, still provide a good fit to the simulated HMF in DE cosmologies at any redshift, provided the correct growth rate of linear density perturbations is used in the fit. This result indicated that the universality of the HMF is broadly preserved in homogeneous DE models at least at the $\sim 20\%$ level, and that the differences in the abundance of CDM halos at high $z$ in DE cosmologies are fully captured
by the different linear growth factors. A similar study was also performed by \cite{Lokas_Bode_Hoffman_2004} restricting to the case of a constant equation of state $w\neq -1$, finding results in general agreement with these earlier claims. 

A more detailed investigation of the HMF in homogeneous DE cosmologies has been carried out much more recently by e.g. \cite{Courtin_etal_2011}
with higher-resolution simulations, finding evidence of deviations of the HMF from a universal behavior at the level of about $10\%$ for the case of a Quintessence model with an inverse power-law potential. The comparison of these results already shows how the improvements in the simulations accuracy and dynamical range have allowed to detect progressively finer details of the imprint of DE on structure formation processes.\\

The effects of a homogeneous DE field on the internal properties of cluster-size halos was then studied in much finer details in \cite{Dolag_etal_2003} by running high-resolution zoom re-simulations of the 17 most massive halos identified in a fiducial $\Lambda $CDM cosmological run within a range of different homogeneous DE cosmologies. For the DE models they considered a constant-$w$ cosmology with $w=-0.6$ and two variable-$w$ models corresponding to Quintessence scenarios with inverse power-law and SUGRA potentials, both with the same value of the equation of state parameter at $z=0$, $w_{0}=-0.86$, and still assuming the same $\sigma _{8}$ normalization of all the models at $z=0$. With such setup \citeauthor{Dolag_etal_2003} investigated the variation of the concentration parameter $c\equiv r_{vir}/r_{s}$ (where $r_{vir}$ is the halo virial radius) in their high-resolution halo sample within the different cosmological models, finding that the overconcentration of halos in DE cosmologies at $z=0$ already highlighted in the early results of \cite{Klypin_etal_2003} can be related to the different linear growth factors through a simple scaling relation given by
\begin{equation}
\label{concentration_scaling}
c_{0}^{DE}=c_{0}^{\Lambda {\rm CDM}}\cdot \frac{D_{+}^{DE}(z_{\rm coll})}{D_{+}^{\Lambda {\rm CDM}}(z_{\rm coll})}\,,
\end{equation}
where $c_{0}$ is the concentration parameter at $z=0$ for a $10^{14}$ M$_{\odot }/h$ halo, $D_{+}(z)$ is the growth factor, and $z_{coll}$ is the collapse redshift of the halo. The same study also showed that the mass dependence of the concentration-Mass relation is not significantly affected in homogeneous DE models, which allows to derive the concentration parameter at $z=0$ within DE cosmologies at any halo mass using Eq.~\ref{concentration_scaling} once the concentration-Mass relation is sufficiently tightly calibrated for the standard $\Lambda $CDM case. 

As a follow-up of this study, \cite{Meneghetti_etal_2004a,Meneghetti_etal_2004b} studied the strong lensing efficiency of the 17 clusters simulated by \citeauthor{Dolag_etal_2003} by means of ray-tracing techniques, finding that the higher concentration of clusters in the DE cosmologies determines a higher lensing efficiency as compared to $\Lambda $CDM, although this effect also crucially depends on the choice for the normalization of the linear matter power spectrum. In fact, a different normalization choice -- assuming e.g. a common amplitude of density perturbations at the last-scattering surface $z_{ls}\approx 1100$ -- would result in the opposite trend for all the main effects of DE cosmologies discussed so far, including a lower halo concentration at $z=0$ as compared to $\Lambda $CDM, and correspondingly a lower efficiency of clusters as strong gravitational lenses. {A similar study was also performed soon after by \cite{Maccio_2005} making use of the simulations of \cite{Klypin_etal_2003}, leading to consistent results with the earlier study of \citeauthor{Meneghetti_etal_2004a}}\\

\begin{figure*}
\includegraphics[width=0.6\linewidth]{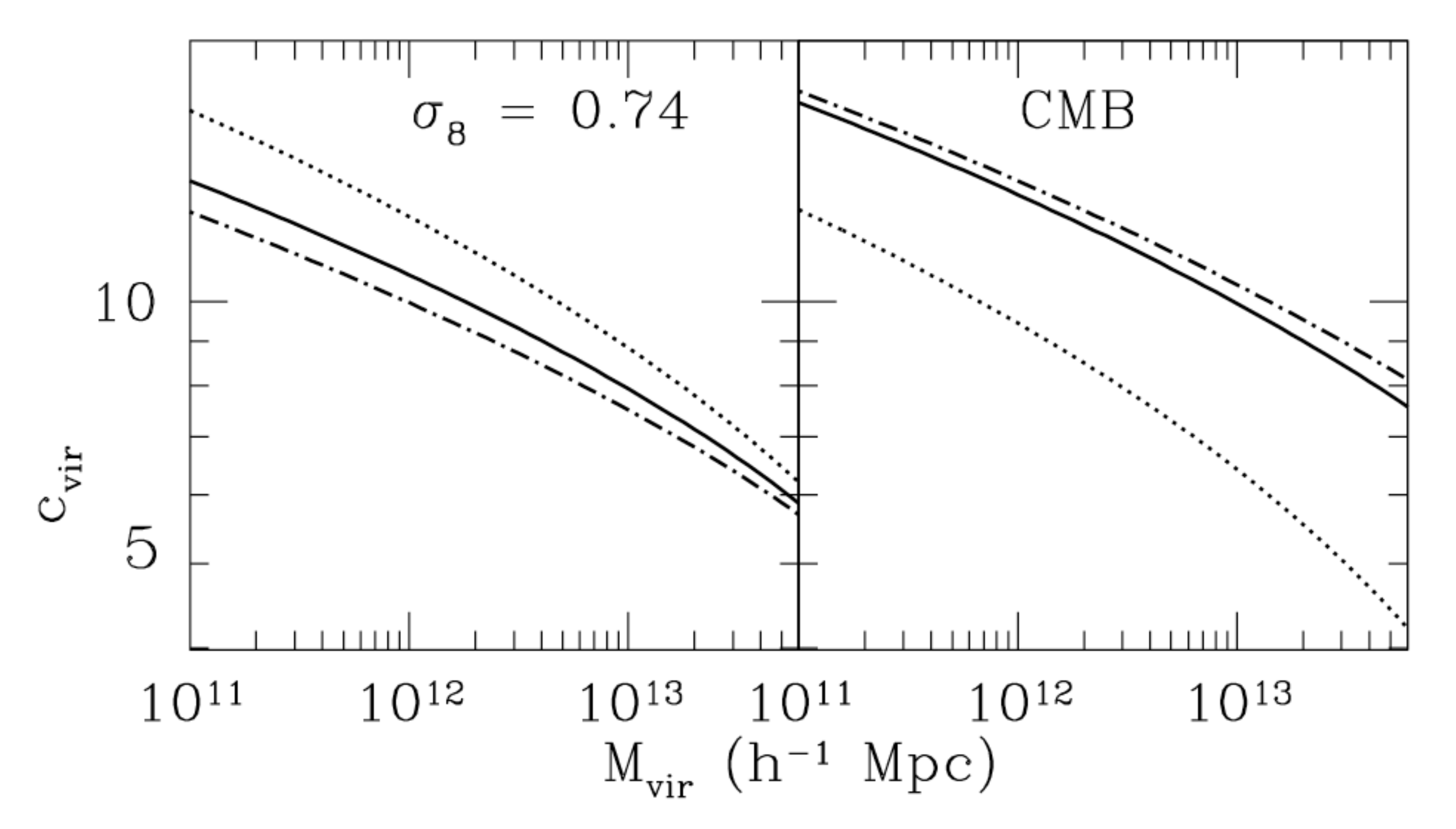}
\includegraphics[width=0.4\linewidth]{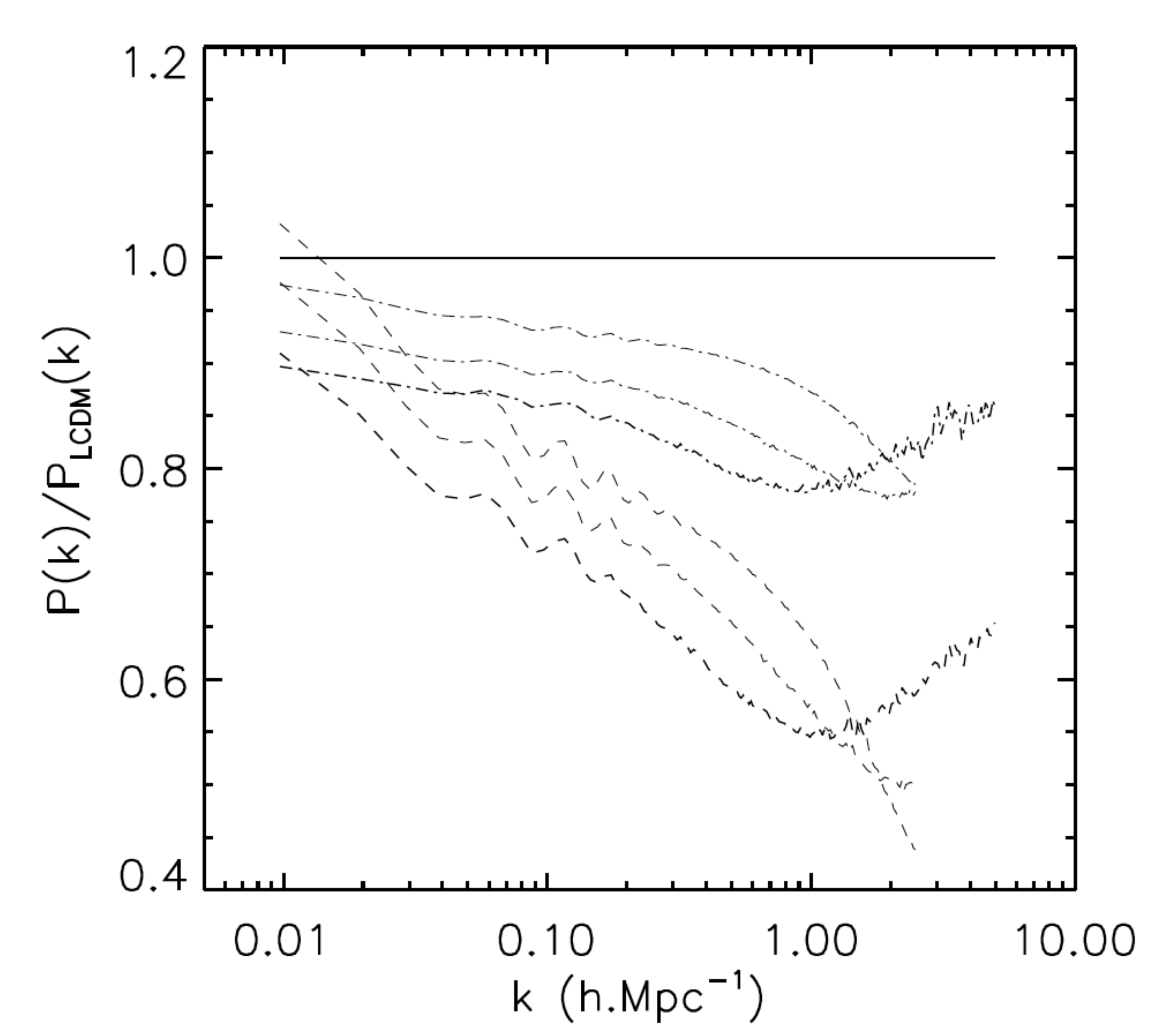}
\caption{{\em Left, plot from \cite{Kuhlen_etal_2005}} - The concentration-mass relation at $z=0$ as extracted from a series of N-body simulations of DE cosmologies with $w\neq -1$, including also the case of ``Phantom" DE, $w<-1$. The plot shows the crucial role played by the normalization choice of the linear perturbations amplitude: on the left, the concentration-mass relations for a  $w=-0.5$ DE model (dotted) and a $w=-1.5$ ``Phantom" DE model (dashed) show that these types of DE scenarios determine respectively an increase and a decrease of the average halo concentration as compared to a standard $\Lambda $CDM cosmology (solid). On the right, the same plot shows the opposite trend for simulations where a common normalization of linear density perturbations at last scattering has been chosen. {\em Right, plot from \cite{Alimi_etal_2010}}: The ratio of the nonlinear matter power spectrum in different DE cosmologies over the standard $\Lambda $CDM case, as extracted from simulations with different final values of the linear perturbations amplitude. The dashed and dot-dashed curves represent the case of Quintessence cosmologies with an inverse power-law and a SUGRA potential, respectively. Different curves refer to different epochs, from top to bottom $a=0.3\,,0.5\,,1$. The plot shows that the maximum deviation from $\Lambda $CDM is obtained at intermediate scales $k\sim 1\, h/$Mpc, and for low redshifts.}
\label{fig2}
\end{figure*}

The issue of the power spectrum normalization was discussed also in \cite{Kuhlen_etal_2005}, that investigated a series of DE models with constant
equation of state $w\neq -1$, extending for the first time the analysis to the case of $w<-1$, generally indicated with the term ``Phantom" DE. Besides showing that an equation of state parameter more negative than the $\Lambda $CDM value with a common normalization of the linear power spectrum to the same $\sigma _{8}$ at $z=0$ determines the opposite trend in the resulting HMF and halo concentration as compared to the $w>-1$ case, this study also explicitly showed that such trends are in any case reversed if one assumes a common normalization of all the cosmologies to the amplitude of scalar perturbations at last-scattering (see Fig.~\ref{fig2}, left). This result confirms and significantly reinforces the early conclusion that
the nonlinear effects of homogeneous DE cosmologies as compared to $\Lambda $CDM are mainly driven by the different evolution with redshift of the linear perturbations amplitude in the DE cosmologies due to their different growth factors $D_{+}(z)$.\\

{The evolution of the baryonic component of the Universe within N-body simulations should be treated taking into account the collisional nature of baryons as opposed to the collisionless nature of CDM particles. Therefore, a variety of methods have been developed to solve the hydrodynamical equations of a perfect gas fluid, along with the solution of the gravitational interaction of masses \citep[see e.g.][]{Teyssier_2001,Springel_2010,Springel_2011}. Additionally, a number of non-adiabatic astrophysical processes can be included in the simulations, ranging from the radiative cooling of the gas and the following formation of stars, to the feedback provided by supernovae explosion and/or by the accretion of gas onto supermassive back holes, to the interaction between the gas and large-scale magnetic fields. The former and simpler approach generally goes under the name of {\em adiabatic} or {\em non-radiative} hydrodynamical simulations, while the latter is referred to as {\em radiative} hydrodynamics.}

The first attempt to include hydrodynamical processes in cosmological simulations of homogeneous DE cosmologies was performed by \cite{Maio_etal_2006}, who studied the formation of the early gas clouds responsible for the reionization of the Universe in a variety of DE cosmologies, {by means of radiative simulations including gas cooling.}
In this work it was found that the earlier formation of structures that characterizes DE models with $w>-1$ applies also to gas clouds that can then induce an earlier reionization epoch as compared to $\Lambda $CDM, a result that  looked very appealing at the time due to the high value of the reionization redshift derived from the first-year data of the WMAP satellite \citep[][]{wmap-1}. The same study also showed that a running spectral index of the primordial power spectrum $n_{s}(k)$  might significantly mitigate this effect, simply by suppressing the small-scale power with respect to the large-scale normalization of the linear perturbations. 

{Radiative} hydrodynamical cosmological simulations were also employed by several other authors to study the structural properties of galaxy clusters. \cite{Aghanim_daSilva_Nunes_2009} performed a series of hydrodynamical runs with gas cooling for a range of homogeneous DE models with constant and variable $w$ to study the impact of DE on the scaling relations between cluster masses and different mass proxies such as the cluster X-ray luminosity and the Sunyaev-Zeldovich (SZ) signal, finding that homogeneous DE does not significantly alter the standard scaling relations and concluding that the use of standard $\Lambda $CDM scaling relations also for homogeneous DE models seems generally appropriate. 

A similar analysis was performed by \cite{DeBoni_etal_2011,DeBoni_etal_2012}, that studied the concentration-Mass relation, the luminosity-temperature relation, and the baryon fraction of clusters  in hydrodynamical simulations of DE models including gas cooling and star formation, finding results in general agreement with previous claims.\\

The impact of homogeneous DE on the nonlinear matter power spectrum was then investigated in detail by e.g. \cite{Ma_2007,Francis_Lewis_Linder_2007,Casarini_Maccio_Bonometto_2009}. In particular both \citeauthor{Francis_Lewis_Linder_2007} and \citeauthor{Casarini_Maccio_Bonometto_2009} found that the nonlinear matter power spectrum of DE models with a variable equation of state parameter $w(a)$ can be derived from the nonlinear power spectrum of constant-$w$ models with an accuracy down to $\sim 1\%$ through a transformation involving only background quantities. 
\cite{Alimi_etal_2010} then also investigated the nonlinear matter power spectrum in some specific DE models selected to best fit background and linear perturbations observational data. This implies that the $\sigma _{8}$ normalization at $z=0$ of the different models be different, and generally lower, in DE cosmologies as compared to $\Lambda $CDM. With such setup \citeauthor{Alimi_etal_2010} found that the maximum deviation of the DE power spectra with respect to $\Lambda $CDM occurs at intermediate scales around $k\sim 1\, h/$Mpc (see Fig.~\ref{fig2}, right panel). Such behavior has been subsequently broadly confirmed also by \cite{Fedeli_Dolag_Moscardini_2011} for different choices of the homogeneous DE evolution with a similar setup, by means of cosmological simulations including also radiative hydrodynamical processes as gas cooling and star formation. \citeauthor{Fedeli_Dolag_Moscardini_2011} additionally showed that star formation efficiency is generally reduced in DE cosmologies \citep[consistently with the earlier results of][]{DeBoni_etal_2011}. 

This maximum deviation from the $\Lambda $CDM matter power spectrum at intermediate scales {(with an amplitude up to 40\% for the most extreme model considered by \citeauthor{Alimi_etal_2010})} appears to be {mostly driven by} the different $\sigma _{8}$ normalization of the various cosmologies that was adopted both by \cite{Alimi_etal_2010} and \cite{Fedeli_Dolag_Moscardini_2011} \citep[a similar feature occurs, for example, also for the case of Coupled DE models with high-$z$ normalization, see e.g.][and the related discussion above]{CoDECS}. {In fact, although} other studies employing a common $\sigma _{8}$ normalization at $z=0$ \citep[such as e.g.][]{Ma_2007} {did also find a qualitatively similar effect, its amplitude results much weaker, with a maximum detected deviation of the order of a few percent}. 
{However, such small residual deviation in the nonlinear matter power spectrum found for simulations with the same $\sigma_{8}$ normalization represents a very important result as it demonstrates how the full nonlinear matter power spectrum cannot be uniquely determined with arbitrary precision by the amplitude and shape of the linear one. More specifically, this result shows that two cosmological models with the same normalization of the linear matter power spectrum at the present epoch but 
with different growth histories can in principle be distinguished from each other through their nonlinear power spectra, although the deviation is expected to be small and consequently particularly difficult to detect.}\\

\begin{figure*}
\includegraphics[height=2.5in]{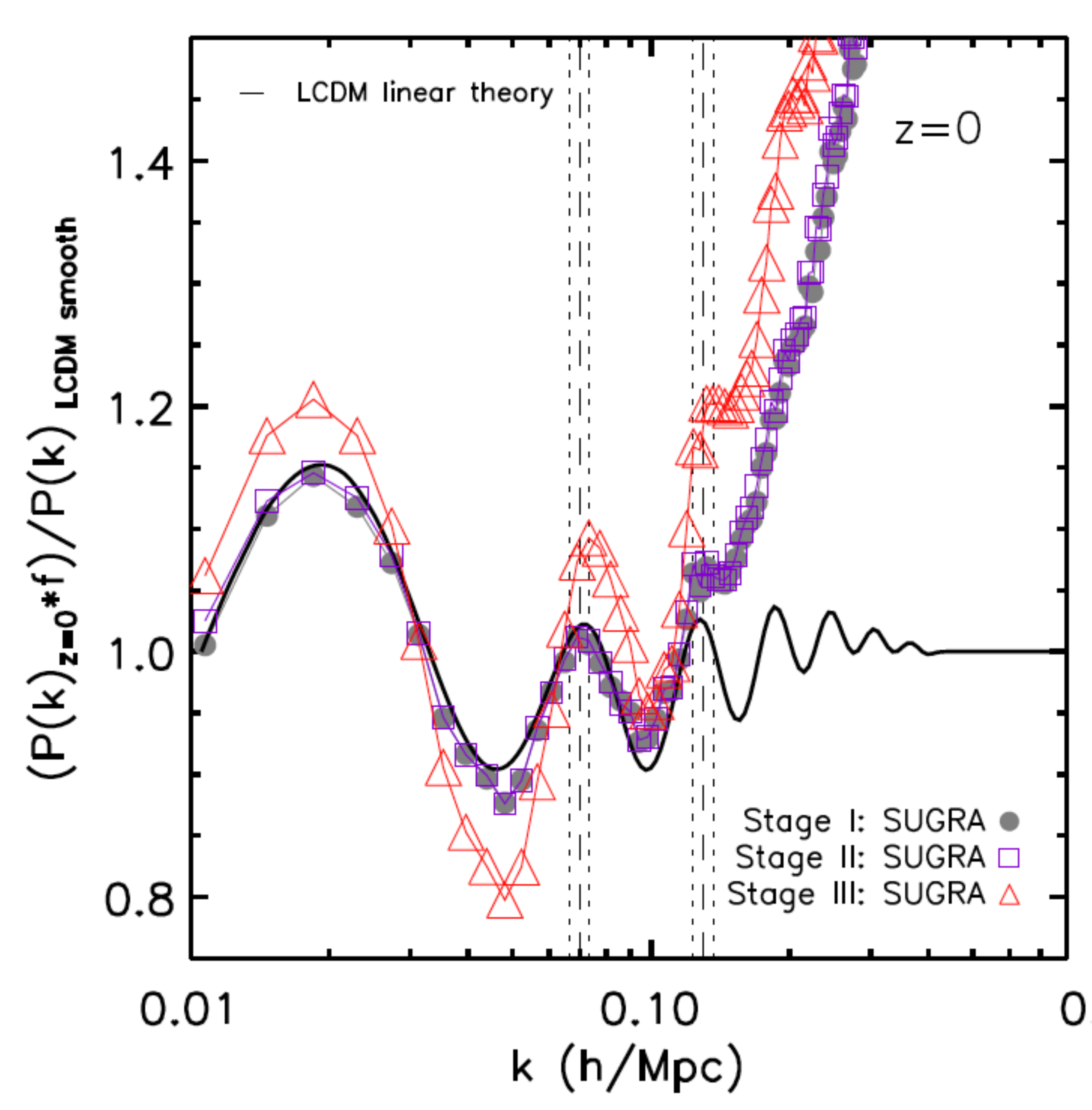}
\includegraphics[height=2.5in]{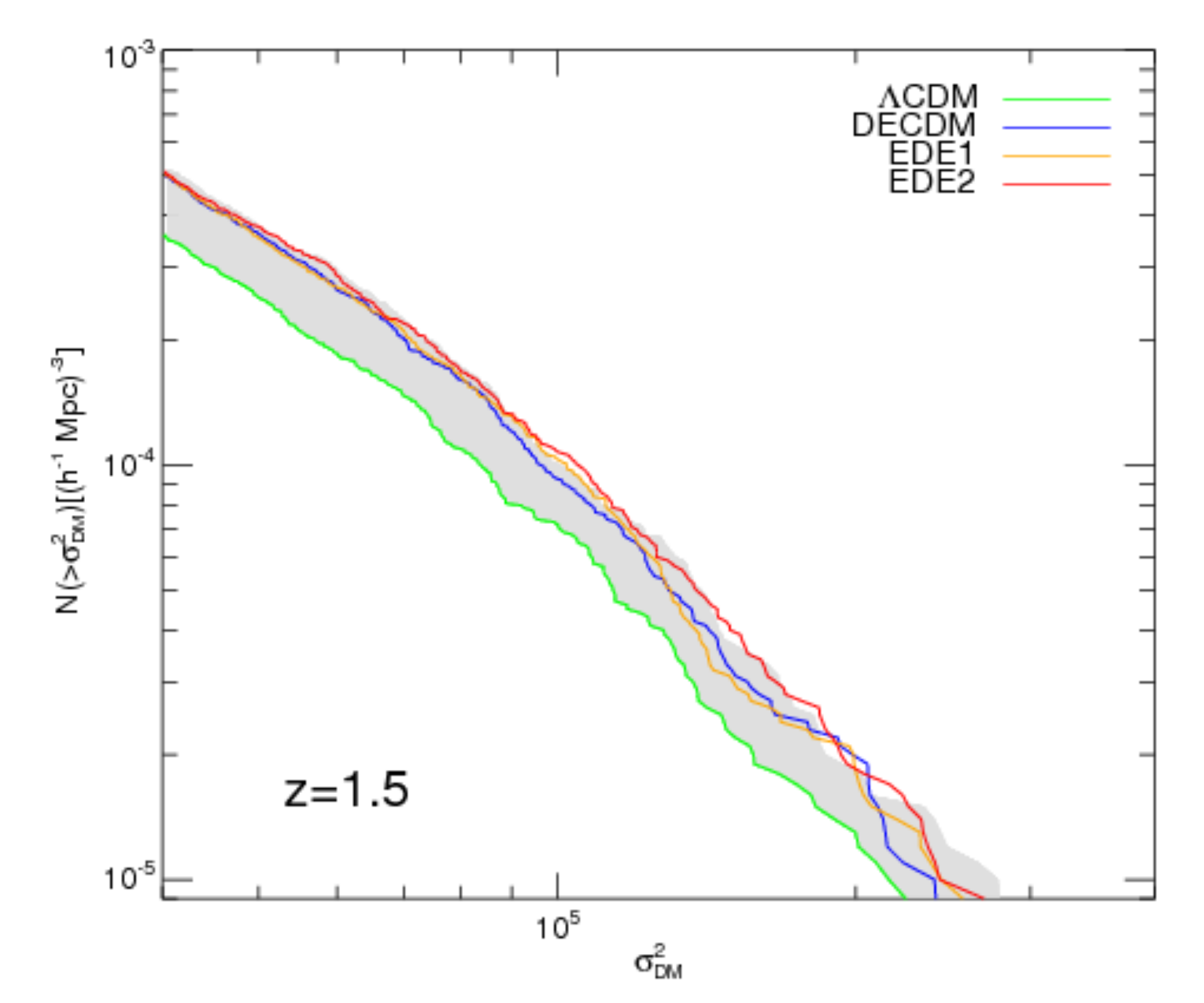}
\caption{{\em Left, plot from \cite{Jennings_etal_2009}}: The nonlinear matter power spectrum divided by a smooth power spectrum (i.e. a power spectrum without baryonic wiggles) for the standard $\Lambda $CDM cosmology (black solid line) and a Quintessence model with a SUGRA potential (red triangles). Although the amplitude of the baryonic acoustic oscillation signal is amplified in the Quintessence model, the location of the peaks coincides with the $\Lambda $CDM case (indicated by the vertical dashed lines) within a $5\%$ accuracy. {\em Right, plot from \cite{Grossi_Springel_2009}}: The velocity function $N(>\sigma )$ as a function of the halo line-of-sight velocity dispersion $\sigma $ as extracted from a $\Lambda $CDM simulation (green line) and a series of DE cosmologies, including two distinct EDE models (orange and red lines). The grey-shaded area corresponds to the gap between the fiducial $\Lambda $CDM cosmology assumed in this work, with $\sigma _{8}=0.8$, and a $\Lambda $CDM model with a higher $\sigma _{8}=0.9$.}
\label{fig3}
\end{figure*}

Another relevant effect of homogeneous DE models on observable quantities that has been investigated through N-body simulations  concerns the Baryon Acoustic Oscillation (BAO) peak in the correlation function of collapsed halos. \cite{Jennings_etal_2009} carried out a series of large CDM-only N-body runs with a modified version of the TreePM code {\small GADGET-2} \citep[][]{gadget-2} within a box of $1500$ Mpc$/h$ aside filled with $646^{3}$ CDM particles for a range of homogeneous DE models with a parametrized equation of state $w(a)$. For this study, \citeauthor{Jennings_etal_2009} adopted a different parameterization with respect to the ones introduced in Section~\ref{sec:models}, following the conclusion  \citep[][]{Bassett_Corasaniti_Kunz_2004} that the CPL parameterization does not reproduce with sufficient accuracy the evolution of Quintessence models at high redshifts. Instead, they employed a four-parameter parameterization proposed by \cite{Corasaniti_Copeland_2003} that provides a better fit to the real equation of state evolution for a wide range of Quintessence cosmologies. The outcomes of the simulations by \citeauthor{Jennings_etal_2009} concerning the matter power spectrum and the HMF in DE models were found to be in good agreement with previous findings. Additionally, this work investigated for the first time the effects of homogeneous DE on the properties of BAOs, finding that even DE models with a significantly rapid evolution of the equation of state parameter at relatively low redshifts do not imprint any significant shift in the location of the BAO peaks as compared to $\Lambda $CDM, thereby making it difficult to detect an evolution of the DE equation of state through measurements of the BAO scale (see Fig.~\ref{fig3}, left panel).\\

The case of the Early Dark Energy (EDE) parameterization of Eq.~\ref{EDE} has been treated separately from Quintessence scenarios and from the CPL parameterization. In two independent and almost contemporaneous
works, \cite{Francis_Lewis_Linder_2008} and \cite{Grossi_Springel_2009} investigated a range of EDE cosmologies by means of CDM-only  simulations performed with two independently-developed modified versions of the N-body code {\small GADGET}, with a particular focus on the impact of EDE on the HMF. Both these works consistently found that the HMF at $z=0$ is only mildly affected by the existence of a non-vanishing fraction of DE at early times, and that the universality of the HMF shape encoded by standard $\Lambda $CDM analytical formulae as e.g. the Jenkins and Warren fitting functions \citep[][]{Jenkins_etal_2001,Warren_etal_2006} is preserved in EDE cosmologies at least at the level of $\sim 10-15\%$ accuracy. These results are in contrast with previous claims by \cite{Bartelmann_Doran_Wetterich_2006} based on a spherical collapse treatment of the formation of CDM halos in EDE cosmologies, which found a significant change in the linear overdensity at collapse $\delta _{c}$ in the presence of an EDE component,  and with the subsequent derivation by \cite{Fedeli_Bartelmann_2007} of a corresponding significant enhancement in the strong lensing efficiency of clusters within EDE cosmologies. Such discrepancy has been further discussed by \cite{Francis_etal_2008b} who showed how under the assumption of small DE perturbations at astrophysical scales (which is the main assumption for homogeneous DE cosmologies and that was implicitly assumed in both the numerical studies mentioned above) a value of the overdensity parameter $\delta _{c}$ close to the standard $\Lambda $CDM value of $\delta _{c}=1.686$ is restored.

The study of \citeauthor{Grossi_Springel_2009} also investigated the concentration-mass relation in the context of EDE cosmologies, and the velocity function $N(>\sigma )$, which is a conceptually similar observable to the HMF where CDM halos are counted based on their line-of-sight velocity dispersion $\sigma $ rather than by their total mass. Such analysis led to the interesting conclusion that the velocity function of EDE cosmologies at redshifts around $z\sim 1.5$ mimics a $\Lambda $CDM velocity function for a standard cosmology with a higher normalization of the linear perturbations amplitude $\sigma _{8}$ (see Fig.~\ref{fig3}, right panel). This result indicates that the presence of an EDE component might be detected through the determination of an excessively large value of $\sigma _{8}$ from the line-of-sight velocity dispersion of high-$z$ clusters.\\

The case of the Vector DE models briefly mentioned in Section~\ref{sec:models} has also been recently investigated with N-body simulations by \cite{Carlesi_etal_2011,Carlesi_etal_2012}. In their 
set of simulations, \citeauthor{Carlesi_etal_2012} investigated the impact of Vector DE models on a number of observables as e.g. the cluster number counts as a function of redshift, 
the HMF, the distribution of cosmic voids, and the structural properties of collapsed halos encoded by the concentration, spin, and shape parameters. As a main conclusion of their analysis, \citeauthor{Carlesi_etal_2012} showed that even though the large-scale properties of structures evolve quite differently in Vector DE cosmologies as compared to $\Lambda $CDM, such deviations are mainly driven by the different evolution of the background cosmological parameters and of the linear growth factor in the different models. On the other hand, as expected, the properties of collapsed structures do not appear to change significantly in Vector DE models, since no direct effect on the gravitational dynamics of particles is present in these cosmologies. Nevertheless, the growth rate of density perturbations shows a very peculiar shape in Vector DE cosmologies that clearly allows to distinguish these models from standard Quintessence scenarios.\\

The study of homogeneous DE models by means of their effects on nonlinear structure formation is presently entering the challenging era of precision cosmology, with a wealth of high-quality data expected for the near future. 
{This implies the} need to move from mainly qualitative assessments of the imprints of DE on the statistical and structural properties of self-gravitating systems to highly reliable quantitative estimations of the expected observational footprints of each specific realization of a homogeneous DE field beyond $\Lambda $. Present and future simulations will then need to face the challenge of significanlty reducing statistical uncertainties mainly related to sample variance and to keep under control systematic effects due to numerical inaccuracies and most importantly to the yet poor understanding of sub-grid physical processes that are expected to heavily affect observable quantities at small scales \citep[see e.g.][]{Semboloni_etal_2011}. 

The former issue can be addressed by running larger cosmological simulations in terms of periodic box size, provided a sufficient mass resolution to resolve collapsed halos over a large enough range of masses can be achieved. Some attempts in this direction are presently being pursued with the Dark Energy Universe Simulations Series \citep[DEUSS,][]{Rasera_etal_2010, Alimi_etal_2012} that aims to perform CDM-only simulations for the fiducial $\Lambda $CDM cosmology and for a few selected homogeneous DE models over simulated volumes comparable with the full observable Universe, employing a modified version of the AMR N-body code {\small RAMSES} \citep[][]{Teyssier_2001}. Such a challenge clearly requires highly sophisticated numerical tools with extremely high scalability and a dedicated pipeline for on-the-fly data compression to maintain the volume of processed data still manageable. 

The latter issue, instead, does not show a similarly clear path towards possible solutions, and significant efforts will have to be made in the near future to refine our understanding of baryonic physics and astrophysical processes playing a substantial role in shaping the properties of cosmic structures at small scales, before these will be readily usable for cosmological studies.

\section{Simulating Dark Energy perturbations and interactions}
\label{sec:inhomogeneous}

As it was discussed in Section~\ref{sec:models}, if the assumption of homogeneity of the DE field at sub-horizon
scales is dropped, the effects of DE on the evolution of density perturbations and on the formation of linear and nonlinear structures in the Universe are not confined anymore only in the Hubble friction term of Eq.~\ref{perturbations1}, but can arise also through a direct contribution of the DE perturbations to the peculiar gravitational potentials experienced by matter particles, as in Eq.~\ref{perturbations2}, or even through additional interactions directly mediated by the inhomogeneous DE degree of freedom, as in Eqs.~\ref{fifth-force},\ref{fifth-force-fric}. 

Such scenarios require much more sophisticated algorithms to be properly implemented in N-body codes as compared to the simpler case of a homogeneous DE field, which only requires to account for a modified expansion history through the correct Hubble function $H(a)$. In the most general case, one should in fact devise algorithms capable to accurately solve a nonlinear Poisson equation like Eq.~\ref{modified_poisson} for an arbitrary matter distribution with periodic boundary conditions. This is an extremely challenging task, and the attempts to include such a sophisticated solver into N-body codes will be reviewed towards the end of this Section. However, this effort is in many cases not strictly necessary, as several specific DE models, although featuring sub-horizon perturbations and/or additional interactions, provide ways to directly relate the scalar field perturbations $\delta \phi $ to the matter distribution in the simulation box through simplified linear differential equations or even through algebraic relations. In this Section, I will review the results obtained with N-body simulations for these different classes of DE scenarios.\\

\subsection{Non-universal couplings}

As no simulations have been performed so far for the case of non-interacting inhomogeneous DE cosmologies, as for instance the Clustering DE scenario introduced in Section \ref{sec:models} \citep[see e.g.][]{Sefusatti_Vernizzi_2011}, I will directly move to review the results obtained in the last years for interacting DE models with non-universal couplings. These are scenarios for which explicit screening mechanisms at small scales are not strictly necessary, which allows to significantly simplify the relation between the DE-mediated fifth-force and the matter distribution. A vast literature is available for a thorough description of the main features of this kind of interacting DE models, see e.g. \cite{Amendola_2000,Amendola_2004,Farrar2004,Pettorino_Baccigalupi_2008,CalderaCabral_2009,Koyama_etal_2009,Baldi_2011a}. As discussed above, for such models the scalar field perturbations can be simply related to the standard gravitational potential $\Phi $ through an algebraic proportionality depending only on the coupling function.\\

\begin{figure*}
\includegraphics[width=2.7in]{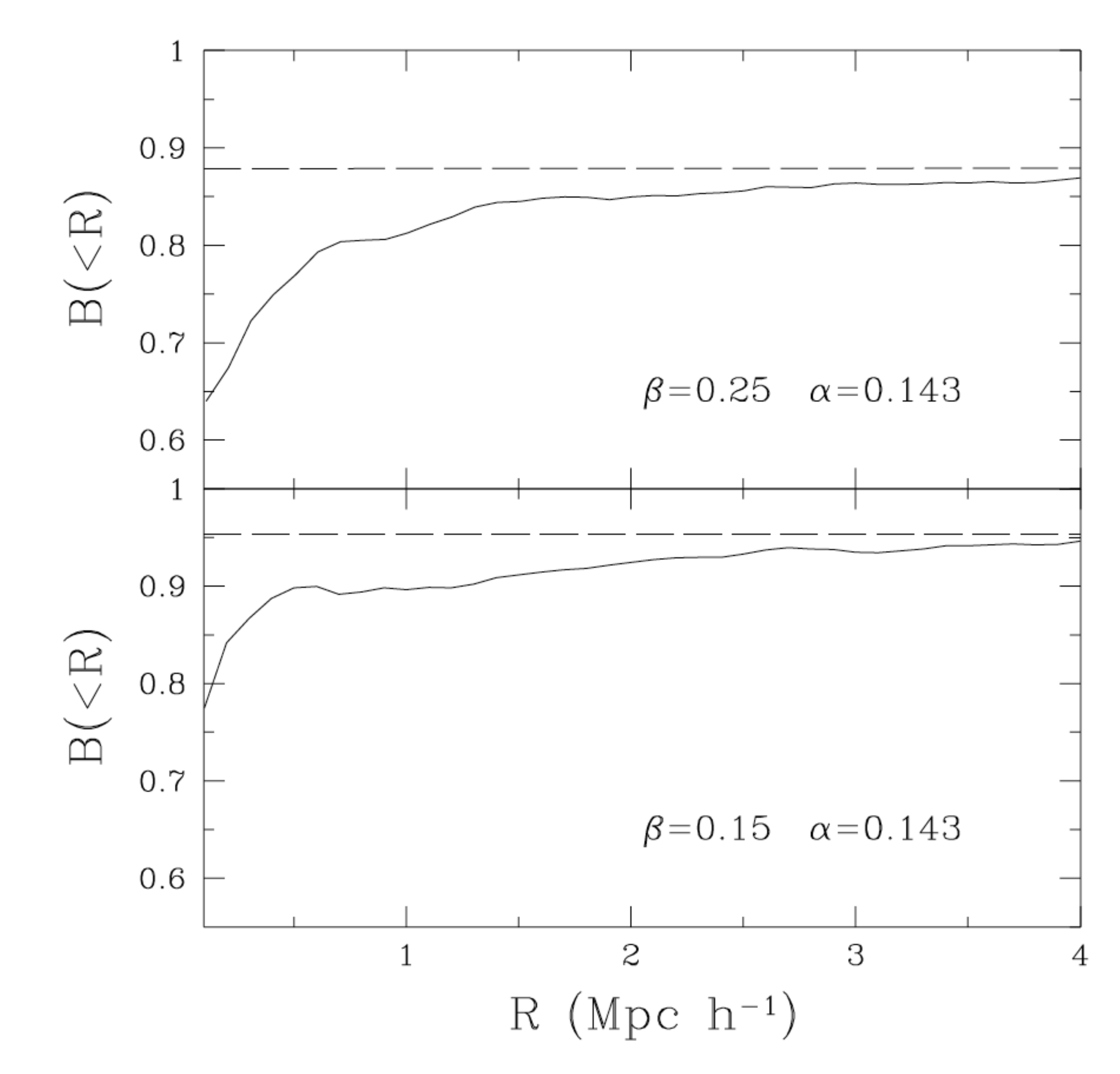}
\includegraphics[width=2.7in]{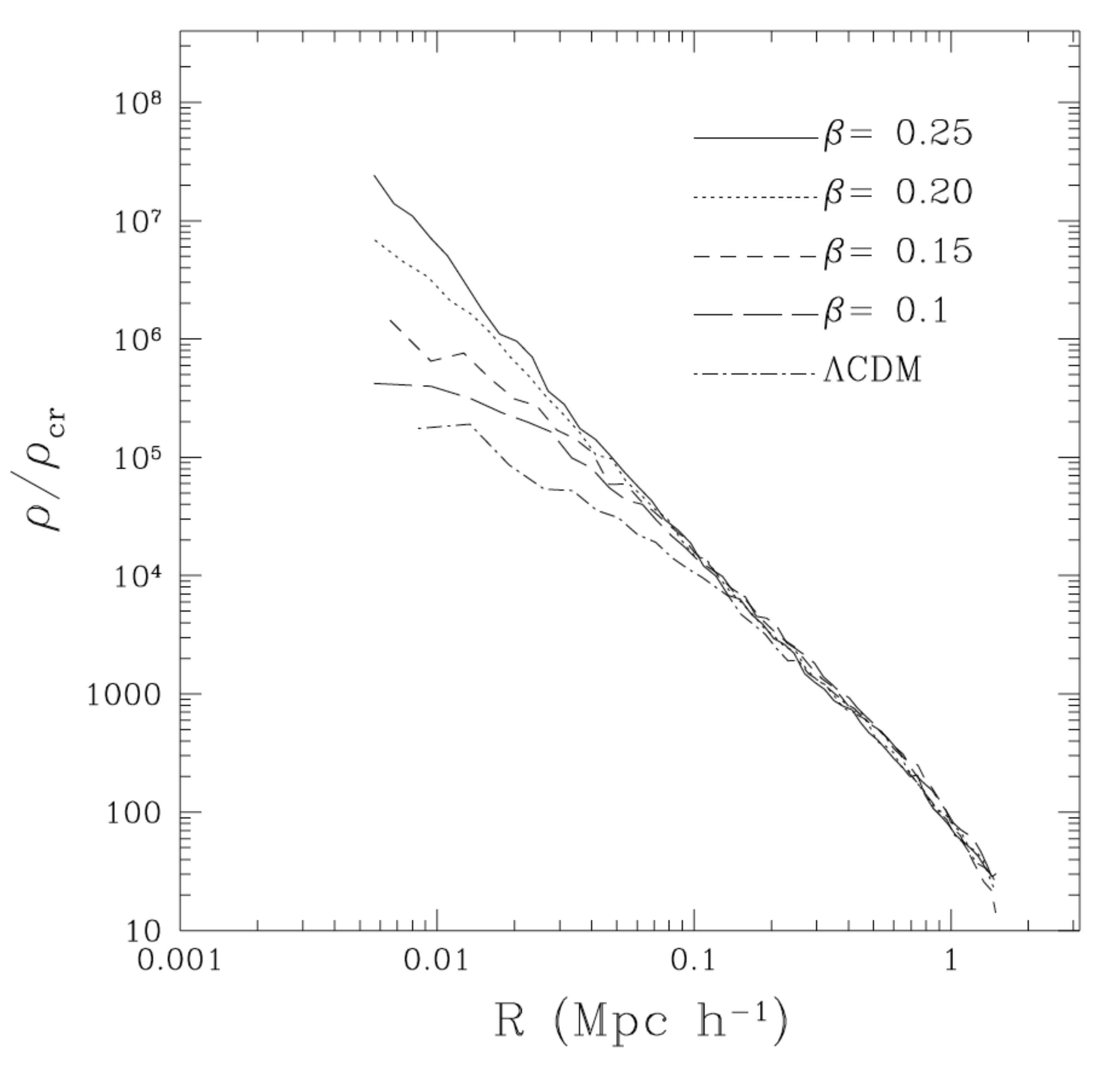}
\caption{{\em Plots from \cite{Maccio_etal_2004}} - {\em Left}: The ratio $B(<R)$ between the baryonic and CDM overdensity enclosed in a sphere of radius $R$ around the center of a massive halo extracted from the N-body simulations of different Coupled DE cosmologies, as a function of radius $R$. The linear bias between baryons and CDM that is visible at large distances from the center, due to the different gravitational dynamics of these two components, is significantly enhanced by nonlinearities in the inner part of the halo. {\em Right}: The density profile of a selected CDM halo forming in the standard $\Lambda $CDM cosmology (dot-dashed) and in various Coupled DE models with different coupling strengths. The effect of the DE-CDM interaction on the density profile appears dramatic in this work, with the maximum coupling $\beta =0.25$ giving rise to a steep power-law behavior of the profile with a logarithmic slope of $\sim -2.3$. Such trend has not been confirmed by the subsequent works of \cite{Baldi_etal_2010} and \cite{Li_Barrow_2011} (see Fig.~\ref{fig5} below).}
\label{fig4}
\end{figure*}

The first N-body simulations of Coupled DE models have been performed by \cite{Maccio_etal_2004} using a modified version of the AMR code {\small ART} for a range of cosmological models based on a DE scalar field
with an inverse power-law potential of the form of Eq.~\ref{RP} interacting with CDM only (i.e. with a vanishing coupling to baryons $\beta _{b}=0$) through a constant coupling function $\beta _{c}$ in the range $0-0.25$. 
All the models were normalized to have the same amplitude of linear density perturbations at the present epoch, and evolved with a self-consistent background expansion history $H(a)$.
The early results of \citeauthor{Maccio_etal_2004} showed that the fifth-force acting between CDM particles induces a bias between the amplitude of baryons and CDM perturbations, which is retained and amplified by nonlinear collapsed objects that show a reduced baryon content as compared to the standard $\Lambda $CDM case (see Fig.~\ref{fig4}, left panel), and that the HMF at $z=0$ in Coupled DE models is practically indistinguishable from the $\Lambda $CDM case for a common $\sigma _{8}$ normalization at the present epoch and can be accurately fit by the standard \cite{Jenkins_etal_2001} fitting formula.

Another significant result of the early work of \citeauthor{Maccio_etal_2004} is the dramatic impact that the fifth-force was found to have on the inner slope of the halo density profiles -- and consequently on the normalization of the concentration-mass relation -- for the halos identified in the sample of their simulations: Coupled DE models were in fact found to produce highly overconcentrated halos as compared to $\Lambda $CDM, with a density profile approaching a power-law (and therefore not fit anymore by an NFW shape) with an inner logarithmic slope as low as $\sim -2.3$ for the largest coupling value $\beta _{c}=0.25$ considered in their work (see Fig.~\ref{fig4}, right panel). Such result, which would have determined extremely tight constraints on the DE-CDM coupling as it would significantly worsen the cusp-core tension existing between numerical predictions and observations for the standard $\Lambda $CDM cosmology, was however not confirmed by later independent studies. 

\begin{figure*}
\includegraphics[width=2.7in]{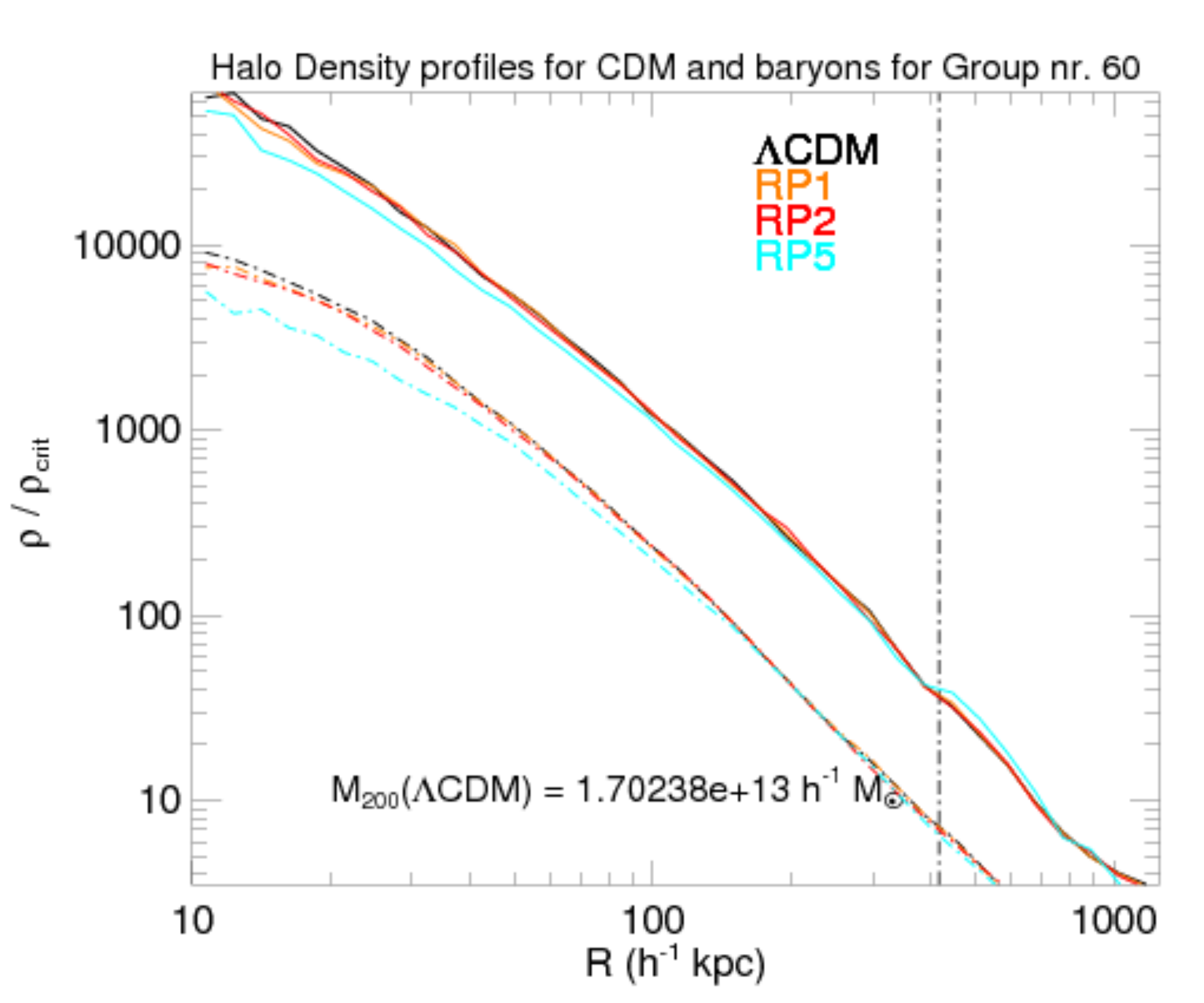}
\includegraphics[width=2.7in]{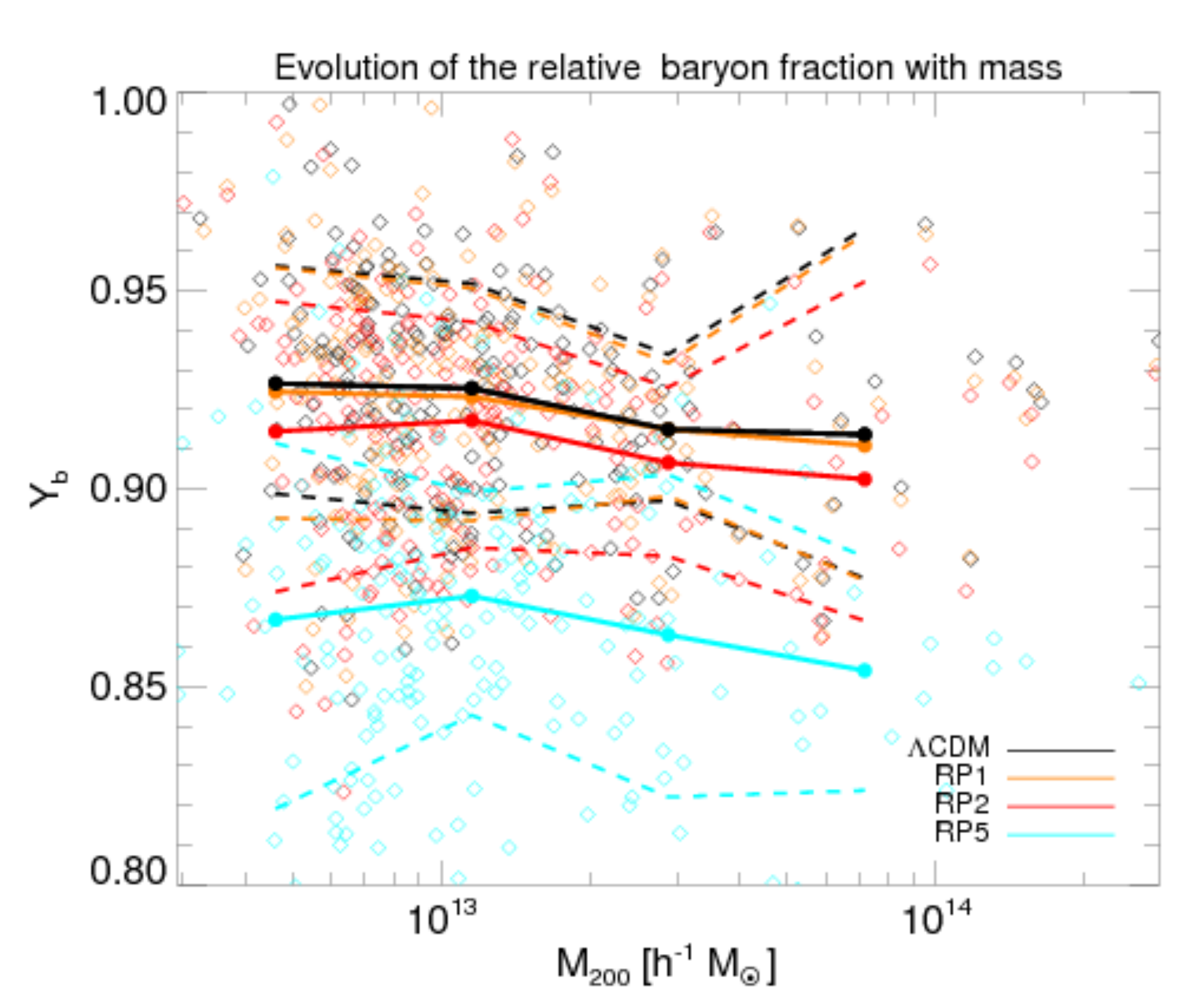}
\caption{{\em Plots from \cite{Baldi_etal_2010}} - {\em Left}: The density profile of baryons (dot-dashed) and CDM (solid) for a massive halo extracted from the N-body simulations of the same Coupled DE models previously investigated by \cite{Maccio_etal_2004} (see Fig.~\ref{fig4} above). The trend of the inner overdensity of the halos as a function of the coupling $\beta _{c}$ found in this work is the opposite of what previously claimed, with Coupled DE models giving rise to a reduction of halo concentrations. {\em Right}: The individual (open diamonds) and average (solid lines) halo baryon fraction in units of the cosmological baryon fraction for a sample including the 200 most massive halos detected in cosmological N-body simulations of the standard $\Lambda $CDM model and of various Coupled DE scenarios. Coupled DE induces a significant reduction of the halo baryon fraction at all masses.}
\label{fig5}
\end{figure*}

In particular, the first {adiabatic} hydrodynamical simulations of Coupled DE cosmologies by \cite{Baldi_etal_2010} -- performed with a modified version of {\small GADGET} -- found essentially the opposite result for the same set of cosmological scenarios: a mild reduction of the inner overdensity of halos for increasing values of the DE-CDM coupling $\beta _{c}$ (see Fig.~\ref{fig5}, left panel), with a consequent systematic shift of the normalization of the concentration-mass relation towards lower concentrations in Coupled DE as compared to $\Lambda $CDM. \citeauthor{Baldi_etal_2010} investigated further this issue by studying the impact of each individual modification of the standard $\Lambda $CDM dynamics implemented in their code, by running test simulations where each of these specific terms was artificially suppressed \citep[see also][for a systematic study of the different dynamical effects in interacting DE scenarios]{Baldi_2011b}. As a result of this analysis, they concluded that the reduction of halo concentration was primarily determined by the effect of the friction term defined in Eq.~\ref{fifth-force-fric} on the local particles dynamics: for a positive coupling $\beta _{c}>0$ and a positive scalar filed velocity $\dot{\phi }$ (which is what is realized for an inverse power-law runaway potential as the one assumed both by \citeauthor{Maccio_etal_2004} and \citeauthor{Baldi_etal_2010}) the friction term $\beta _{c}\dot{\phi }\vec{v}$ acts as an effective drag (i.e. an ``anti-friction") accelerating particles along the direction of their motion. {This corresponds to an injection of} kinetic energy in virialized collapsed systems promoting the migration of particles from inner to outer orbits, thereby adiabatically changing the virial equilibrium of the system towards more extended configurations of the halo core. This general result was then confirmed some time later by the independent collisionless simulations of \cite{Li_Barrow_2011} performed with a modified version of the AMR code {\small MLAPM}, that found a comparable shallowing of the inner density profile of CDM halos in Coupled DE models as the earlier results of \citeauthor{Baldi_etal_2010}

Besides the impact on the inner structure and concentration of collapsed halos, the work of \citeauthor{Baldi_etal_2010} also investigated the specific baryon fraction in massive structures, finding that halos in Coupled DE cosmologies tend to have a significantly lower baryonic content than their $\Lambda $CDM counterparts (see Fig.~\ref{fig5}, right panel), in good agreement with previous results. {Finally,  \citeauthor{Baldi_etal_2010}} studied in detail the evolution with redshift of the HMF, showing that both the analytical expression of \cite{Sheth_Tormen_1999} and the standard fitting function of \cite{Jenkins_etal_2001} reproduce with reasonable accuracy the HMF of Coupled DE cosmologies up to $z\sim 2.5$, provided the correct growth factor $D_{+}(z)$ of each specific model is used for computing the theoretical halo abundance.

The effects of Coupled DE models with a constant coupling to CDM on the high-$z$ intergalactic medium, and in particular on the transmitted Lyman-$\alpha $ flux, has then been studied soon after with a series of {radiative} hydrodynamical N-body simulations by \cite{Baldi_Viel_2010}, allowing to place new independent constraints on the coupling value of about $\beta _{c}\lesssim 0.15$ at $2\sigma $ confidence level, while the impact on the correlation between CDM and galaxy distributions in clusters has been discussed in \cite{Baldi_Lee_Maccio_2011}.\\

\begin{figure*}
\includegraphics[width=2.7in]{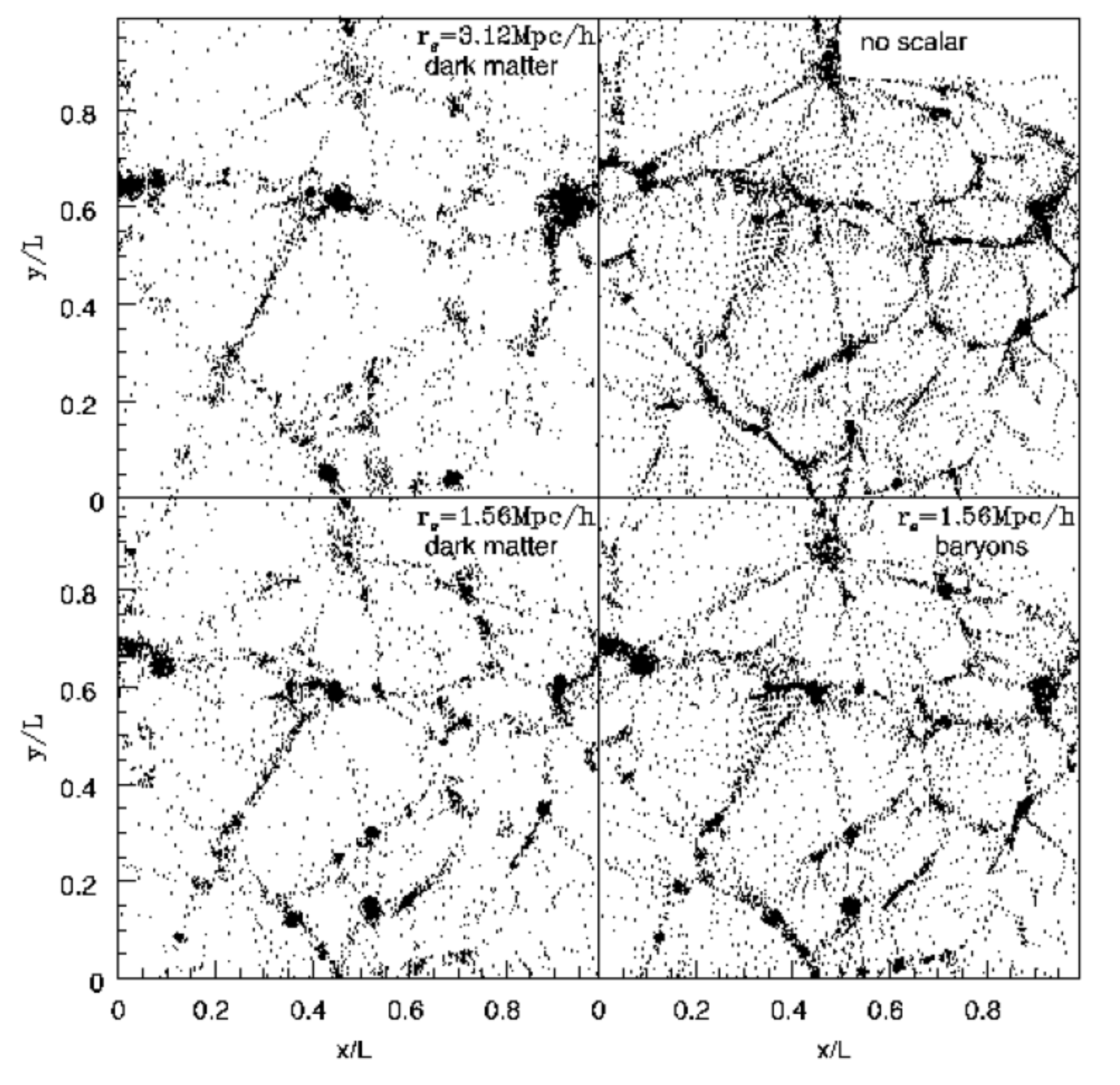}
\includegraphics[width=2.7in]{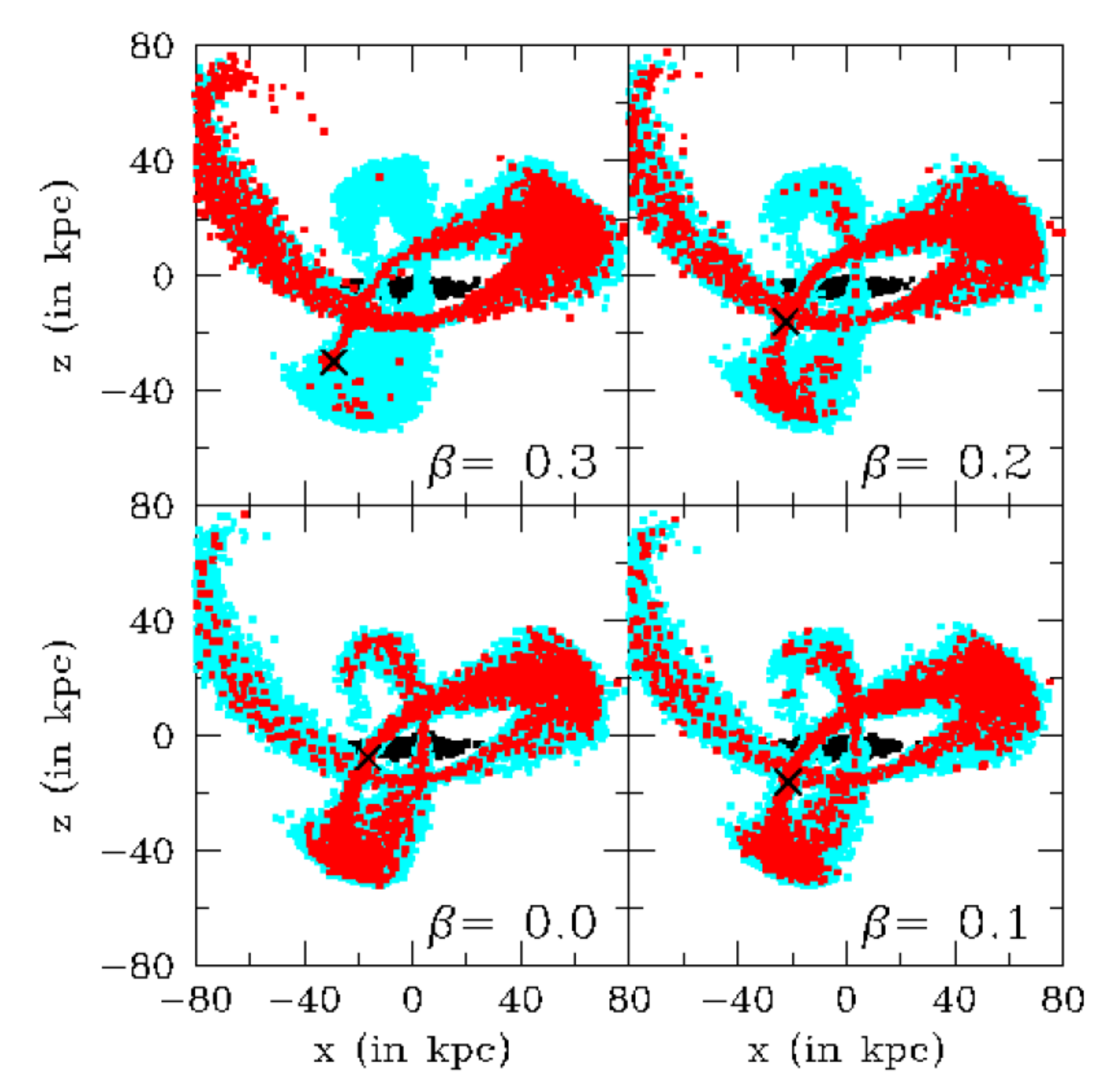}
\caption{{\em Left, plot from \cite{Nusser_Gubser_Peebles_2005}}: the CDM and baryon distribution in a cosmological box of $50$ Mpc$/h$ aside, with and without an additional attractive scalar force with a Yukawa long-range suppression. The right upper corner shows the case of the standard $\Lambda $CDM cosmology, while the other plots display the particle distribution in the presence of a screened fifth-force. The most evident effect appearing from the plots is that cosmic voids are emptier than in the standard cosmological model. {\em Right, plot from \cite{Kesden_Kamionkowski_2006}}: The stellar (red dots) and CDM (cyan dots) components of the remnant streams of a dwarf satellite tidally disrupted by its motion in the gravitational potential of a Milky Way-like spiral galaxy, as extracted from N-body simulations with and without (left lower plot) a scalar fifth-force. In the presence of a fifth-force the leading and trailing streams do no longer appear symmetric in their stellar content, which allows to place constraints on the strength of the fifth-force.}
\label{fig6}
\end{figure*}

A related class of fifth-force models, where however the fifth-force is not necessarily associated with a DE degree of freedom but is rather assumed as a general additional interaction between massive particles, has been investigated by \cite{Nusser_Gubser_Peebles_2005}, who ran cosmological N-body simulations including an additional fifth-force between massive particles, with the further complication that such fifth-force is assumed to be screened at large distances by a Yukawa suppression factor of the form $\exp(-r/r_{s})$ in the fifth-force potential, with $r_{s}$ being a characteristic length scale defining the range of propagation of the scalar fifth-force \citep[see][]{Gubser2004}. In their work, \citeauthor{Nusser_Gubser_Peebles_2005} focused mainly on the effects of the fifth-force on the evolution of cosmic voids, finding that these specific fifth-force scenarios produce a lower CDM and baryon density in voids as compared to $\Lambda $CDM (see Fig.~\ref{fig6}, left panel), which is an appealing feature to address the longstanding problem of dwarf and irregular galaxies within voids being observationally too rare \citep[][]{Peebles_2001}. Similar studies have been subsequently performed also by e.g. \cite{Hellwing_etal_2010,Keselman_Nusser_Peebles_2010}.

The same class of scenarios has then been tested also through N-body simulations of individual galactic-size halos, focusing on the dynamics of dwarf satellites and on the effects of the fifth-force on the details of their tidal remnants. The first work of this kind, performed by \cite{Kesden_Kamionkowski_2006}, found very significant effects of the screened scalar fifth-force on the relative abundance of stars living in the leading and trailing tidal streams of gravitationally stripped dwarf satellites directly comparable to the Sagittarius stream. Such effects allowed \citeauthor{Kesden_Kamionkowski_2006} to put very tight constraints on the maximum allowed value of the scalar interaction. However, a subsequent work by \cite{Keselman_Nusser_Peebles_2009} found significantly different results for different choices of the initial conditions of the system, allowing for significantly larger values of the coupling without conflicting with direct observations of dwarf satellites tidal streams, although a further follow-up paper by \cite{Kesden_2009} challenged in turn the specific initial conditions chosen by \citeauthor{Keselman_Nusser_Peebles_2009}\\

\begin{figure*}
\includegraphics[width=2.7in]{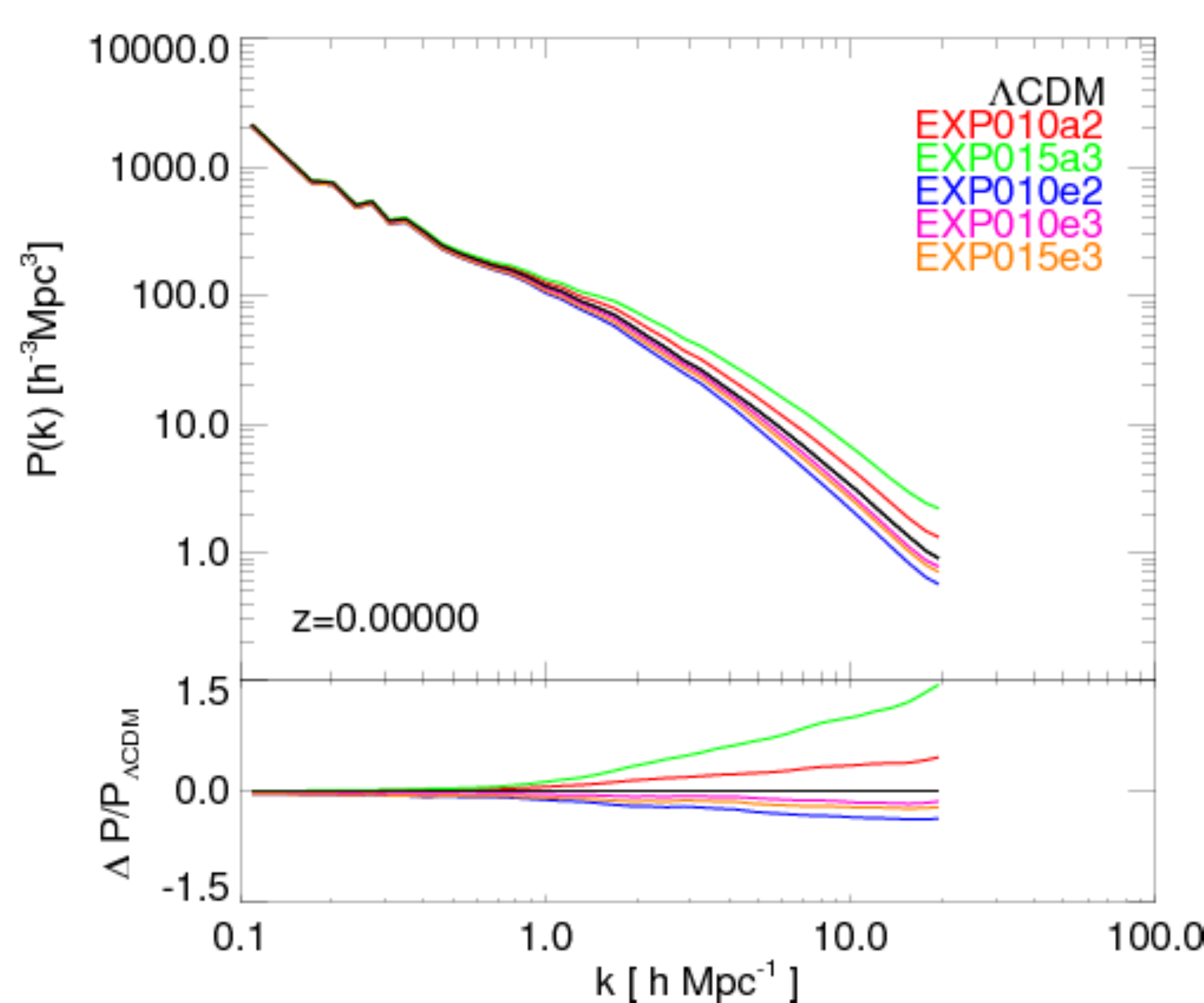}
\includegraphics[width=2.7in]{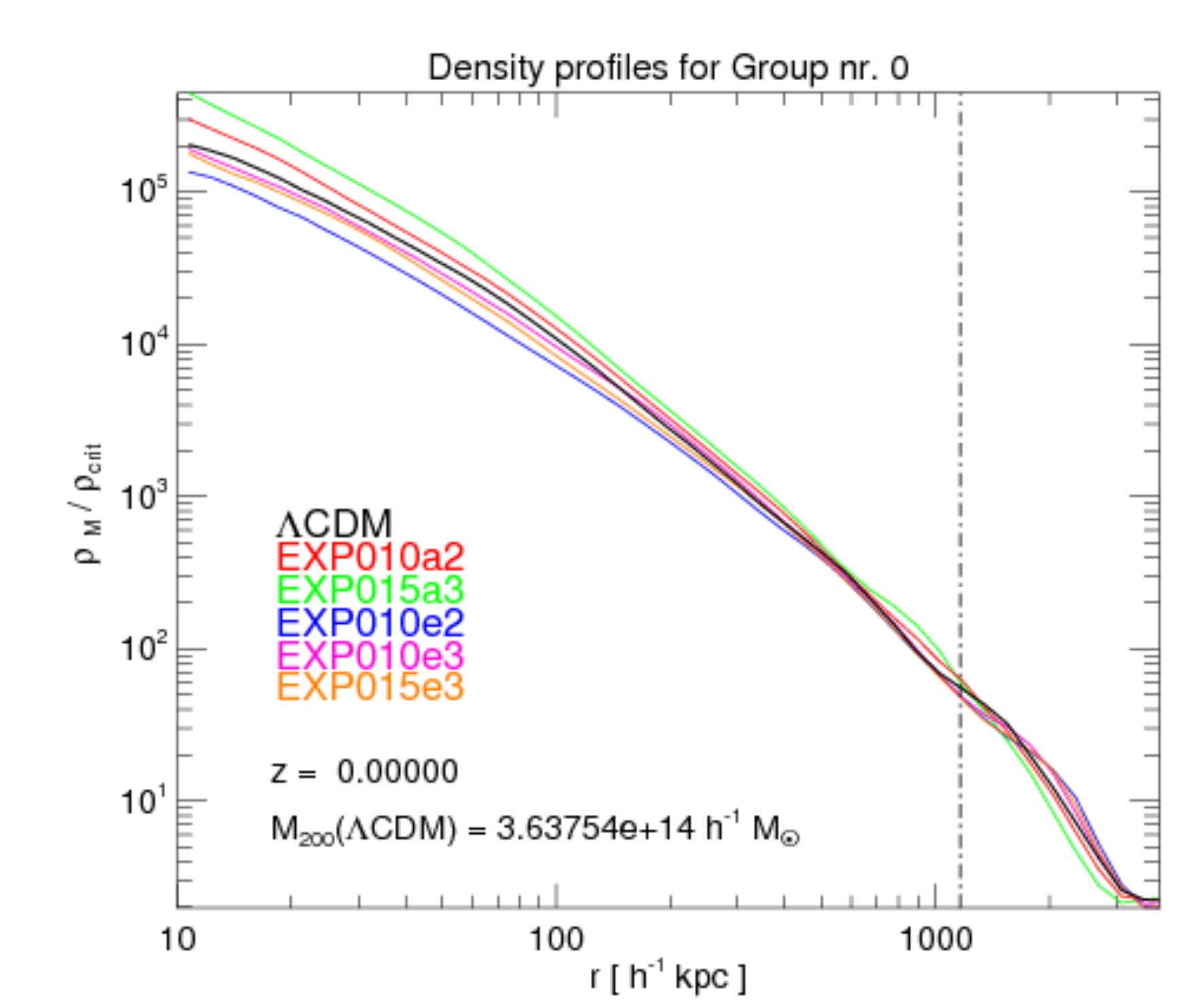}
\caption{{\em Plots taken from \cite{Baldi_2011a}} - The nonlinear matter power spectrum ({\em Left}) and the halo density profile ({\em Right}) as computed from a series of simulations of Coupled DE models with time-dependent coupling functions. 
The time evolution of the coupling induces additional effects on the evolution of structures and can modify the standard $\Lambda $CDM results (black lines) in opposite ways, depending on the specific time evolution of the coupling strength.}
\label{fig7}
\end{figure*}

The possibility of a time-dependent coupling between DE and CDM, representing a more general class of interacting DE cosmologies than the constant coupling models simulated in the early works just discussed, has been included in N-body simulations for the first time in the work by \cite{Baldi_2011a}, that performed a series of {adiabatic} hydrodynamical simulations for different coupling functions including phenomenological parameterizations as e.g. $\beta _{c}(a)\propto a^{\beta _{1}}$ and dynamical evolutions of the coupling such as e.g. $\beta _{c}(a)\propto \exp \left[ \beta _{1}\phi (a)\right] $. In this work, also assuming a common normalization of the different models to the same $\sigma _{8}$ at $z=0$, all the main basic analysis already performed for constant coupling models have been repeated, interestingly showing that the time variation of the interaction induces a whole range of new effects on structure formation processes that are in general absent for the simplified case of a constant coupling. In particular, both the small-scale nonlinear power and the average halo concentrations, which can only be reduced as compared to $\Lambda $CDM within constant-coupling models, can instead show both trends -- i.e. be either reduced or increased -- for variable coupling scenarios, depending on the specific evolution of the coupling function (see Fig.~\ref{fig7}). This is due to the fact that besides the friction term (which as discussed above alters the virial equilibrium of collapsed objects by forcing them to expand) also the time evolution of the effective gravitational constant can modify the virial state of halos, and in particular for a coupling that grows in time this has the effect of favoring more concentrated configurations, thereby counteracting and possibly overcoming the opposite effect of the friction term.\\

The effect of the linear amplitude normalization in interacting DE models has been studied in a series of works using different assumptions for the simulations initial conditions. In particular, \cite{Baldi_Pettorino_2011} and \cite{Baldi_2011c} investigated the effects of Coupled DE models, both with constant and variable coupling functions, on the expected number of massive clusters as a function of redshift, assuming a common normalization of the linear perturbations amplitude in the different models at high-$z$ rather than at $z=0$. These two works have shown how Coupled DE models consistent with the CMB normalization of the 
amplitude of density perturbations at last scattering systematically predict a larger abundance of massive clusters at high redshifts, as a consequence of the additional fifth-force that enhances structure formation {thereby inducing a higher $\sigma _{8}$ normalization at $z=0$ for cosmologies that start from the standard normalization at the redshift of last scattering ($z_{\rm ls}\approx 1100$)}. Such result seems appealing to explain possible detections of extremely massive clusters at high-$z$ that might result difficult to accommodate in the context of the standard $\Lambda $CDM scenario \citep[see e.g.][for an overview on this topic]{Jee_etal_2009,Rosati_etal_2009,Mortonson_Hu_Huterer_2011,Waizmann_Ettori_Moscardini_2011}. In particular, \cite{Baldi_2011c} performed the first N-body simulations of a specific type of Coupled DE models called ``Bouncing" coupled DE, characterized by a constant coupling $\beta _{c}$ and by a SUGRA self-interaction potential for the scalar field $\phi $, resulting in a particular dynamical evolution of the field that allows to match at the same time the normalization of linear density perturbations both at the last scattering surface $z_{ls}\approx 1100$ and at the present time, still allowing for significant deviations from the $\Lambda $CDM behavior at intermediate redshifts. This work has shown that Coupled DE models of the ``Bouncing" type allow to produce a significant excess of massive clusters at high redshifts without overpredicting the cluster counts in the local Universe, contrary to what can be achieved with standard Coupled DE models with constant or variable coupling and even with completely different approaches -- such as e.g. primordial non-gaussianity -- that have been invoked  as a possible explanation for unexpectedly massive high-$z$ clusters \citep[see e.g.][]{Grossi_etal_2007,LoVerde_etal_2008,LoVerde_Smith_2011}.

The different types of Coupled DE scenarios (constant $\beta $, variable $\beta $, ``Bouncing") that have been studied through different (and often not mutually comparable) N-body simulations in the last years, have now been included in a large, systematic, and self-consistent simulations project with the aim to provide accurate and statistically significant numerical data for Coupled DE cosmologies to be readily compared with each other and with standard $\Lambda $CDM predictions. Such initiative goes under the name of the ``{\small CoDECS} Project" \citep[][]{CoDECS} and includes N-body and {adiabatic} hydrodynamical simulations of a variety of cosmological scenarios, all sharing the same WMAP7 \citep[][]{wmap7} cosmological parameters at $z=0$ (except for $\sigma _{8}$) and with a common normalization of the linear density perturbations amplitude at the redshift of the last scattering surface. The post-processed numerical data of the ``{\small CoDECS} Project" (such as nonlinear matter power spectra, halo and sub-halo catalogs, etc...) are made publicly available through a dedicated web-database\footnote{http://www.marcobaldi.it/web/CoDECS} and have already been used for a number of studies aimed at testing Coupled DE models against present or future observational data. In particular, \cite{Lee_Baldi_2011} used these data to investigate the impact of the DE-CDM interaction on the pairwise infall velocity of colliding galaxy clusters morphologically and dynamically comparable to the ``Bullet" cluster \citep[][]{Markevitch_etal_2001}, finding that Coupled DE cosmologies very significantly enhance the probability of high-velocity collisions. \cite{Marulli_Baldi_Moscardini_2012} made use of the {\small CoDECS} public data to investigate the peculiar features of interacting DE in the clustering and redshift-space distortions patterns of galaxies, while \cite{Beynon_etal_2012} computed forecasts for the weak lensing constraining power of the Dark Energy Survey (DES) and the Euclid satellite mission on the DE-CDM interaction, and \cite{Cui_Baldi_Borgani_2012} employed the same data to test the universality of the HMF, finding evidence for deviations from the universal behavior at the level of about $10\%$. \\

A completely different type of models belonging to the class of inhomogeneous and interacting DE cosmologies with a non-universal coupling is given by the ``Growing Neutrino" scenario \citep[proposed by][as a possible solution to the DE coincidence problem]{Amendola_Baldi_Wetterich_2008}, characerized by a constant and very large coupling to massive neutrinos ($|\beta _{\nu} |\gtrsim 50$) while the other matter fields remain uncoupled. The evolution of perturbations in the Growing Neutrino model is characterized by a very fast growth of neutrino density fluctuations soon after the transition from the relativistic to the non-relativistic regime, which  for realistic choices of the model's parameters happens around $z\sim 4-6$ \citep[][]{Mota_etal_2008}. As neutrino perturbations very quickly grow nonlinear, a full N-body treatment is required in order to properly follow the evolution of neutrino structures and of the relative gravitational potential at large scales.

The first N-body simulations of the Growing Neutrino scenario have been performed by \cite{Baldi_etal_2011a} with a suitably modified version of {\small GADGET} for a model with coupling $\beta _{\nu } = -52$ and a neutrino mass at $z=0$ of $m_{\nu ,0}=2.4$ eV. These early simulations have allowed to follow the formation of nonlinear neutrino halos at the expected scales of $\sim 10$ Mpc$/h$ and larger down to $z\sim 1$, and to compute the backreaction effect of the gravitational potential associated with such neutrino lumps on the CDM distribution, showing a clear enhancement of the CDM bulk flow and an excess of CDM power at the largest scales available in the simulation box ($\approx 320$ Mpc$/h$ aside). The limitations of the newtonian approximation, which is generically assumed in N-body solvers, did not allow to run these simulations down to $z=0$ as the strong acceleration experienced by neutrino particles made the neutrinos relativistic again, with velocities comparable and eventually exceeding the speed of light at redshifts lower than $1$. A significant improvement in this respect has been made with the completely independent numerical implementation developed by \cite{Ayaita_Weber_Wetterich_2012} which explicitly includes a fully relativistic treatment of neutrino velocities ensuring that the speed of light limit for particles velocities be automatically fulfilled in the simulation. Additionally, the implementation of \citeauthor{Ayaita_Weber_Wetterich_2012} includes the effects of the local variation of neutrino masses and the backreaction of the formation of neutrino structures on the cosmic background expansion rate that were discarded in the earlier work of \cite{Baldi_etal_2011a}.\\

\subsection{Universal couplings and screening mechanisms}
I now move to consider the case of interacting DE models with universal couplings, i.e. cosmological scenarios where the interaction between an inhomogeneous scalar degree of freedom that can be associated with DE involves all massive particles in the Universe. As discussed above, for such models an explicit screening mechanism capable to suppress the fifth-force in the local neighborhood of the solar system is generally required in order to avoid conflicts with local tests of General Relativity. Nevertheless,  as a first order approximation for models with a coupling strength much smaller than gravity (i.e $\beta ^{2}\ll 1$) and with a sufficiently flat self-interaction potential, the issue of local recovery of standard gravity can be disregarded when focusing on structure formation processes at scales significantly larger than the solar system itself. This is the case of scalar-tensor theories as e.g. Extended Quintessence models \citep[see e.g.][]{Perrotta_etal_2000,Baccigalupi_Matarrese_Perrotta_2000,Pettorino_etal_2005,Pettorino_Baccigalupi_2008} where a cosmic scalar field playing the role of DE directly couples to gravity in the Jordan frame, corresponding to a universal coupling to matter in the Einstein frame \citep[see again][]{Pettorino_Baccigalupi_2008}. Similarly to the case of Coupled DE models, also for Extended Quintessence theories it is then possible to simplify Eq.~\ref{modified_poisson} and directly relate the strength of the extra fifth-force acting (in this case) among all massive particles to the standard gravitational potential through an algebraic equation relating the effective gravitational attraction experienced by massive particles to the standard Newton's constant. However, differently from the case of Coupled DE, in Extended Quintessence models such relation directly depends on the sign of the effective coupling, such that the total interaction between particles can result both stronger or weaker than standard gravity, respectively for positive and negative values of the coupling.\\

The first N-body simulations of Extended Quintessence scenarios have been presented in \cite{DeBoni_etal_2011}. 
They made use of the above-mentioned approximation of the effective fifth-force in terms of the standard gravitational potential for their modified version of {\small GADGET} used to perform a series of {radiative} hydrodynamical cosmological simulations with gas cooling and star formation for a range of DE scenarios including also Extended Quintessence models with both positive and negative couplings. In their simulations, \citeauthor{DeBoni_etal_2011} always assumed a common normalization of linear density perturbations at last-scattering for all the different cosmologies, and focused on the hydrodynamical properties of massive halos corresponding to galaxy clusters. As a main result, \citeauthor{DeBoni_etal_2011} showed that baryonic physics does not appear
to be significantly affected by the additional interaction although the formation history of clusters and consequently their structural properties as well as their past record of star formation are altered by the DE phenomenology. Interestingly, they also found that
both the stellar and gas content of relaxed, massive clusters is not significantly modified in cosmologies where a universal scalar interaction besides gravity is present, as compared to their $\Lambda $CDM counterparts. Such result provides a clear observational way to discriminate between Extended Quintessence scenarios and models with non-universal couplings as the Coupled DE cosmologies discussed above, since for the latter the overall gas content of massive halos significantly changes as a function of time as compared to the standard $\Lambda $CDM case (see Fig.~\ref{fig5}).

The accuracy of the approximation relating the extra fifth-force of Extended Quintessence models to the standard gravitational potential has been explicitly tested by \cite{Li_Mota_Barrow_2011a} making use of a more sophisticated algorithm developed for modified gravity models that feature an explicit screening mechanism (see below) implemented in a modified version of the AMR code {\small MLAPM}. Such algorithm is capable to solve the full nonlinear Poisson equation \ref{modified_poisson} without resorting on any approximation for a wide range of functions $F(\delta \phi )$ -- including the case of scalar-tensor theories like Extended Quintessence -- through a mesh-based iterative relaxation scheme. Therefore, in their simulations \citeauthor{Li_Mota_Barrow_2011a} could explicitly solve for the full scalar field perturbations $\delta \phi (t,\vec{x})$ in a cosmological simulation box with periodic boundary conditions and derive the exact fifth-force acting on each particle within the simulated volume. By directly comparing the exact fifth-force computed with this algorithm to the one obtained by scaling the standard gravitational potential with the approximated relation adopted in \citeauthor{DeBoni_etal_2011}, \citeauthor{Li_Mota_Barrow_2011a}
showed that such approximation is highly accurate even for significantly larger coupling values than the ones investigated by \citeauthor{DeBoni_etal_2011} An explicit solution of the full nonlinear Poisson equation \ref{modified_poisson} is therefore not necessary for Extended Quintessence models. With their simulations \citeauthor{Li_Mota_Barrow_2011a} also showed that several different observables like the nonlinear matter power spectrum, the halo mass function, and the concentration-mass relation are modified in opposite ways as compared to the standard $\Lambda $CDM case depending on the sign of the coupling. This is due to the fact that Extended Quintessence models are characterized by the superposition of two different effects related, respectively, to the modified expansion history and to the extra fifth-force that characterize these cosmologies. In particular, the former effect tends to slow down the growth of linear density perturbations due to a faster expansion rate, while the latter can either enhance or suppress the growth of linear and nonlinear structures due to the larger or smaller effective gravitational constant. As a result, these two effects can either partially balance each other (for the case of a positive coupling, i.e. an enhanced effective gravity) or conspire towards a significantly slower growth of perturbations (for the case of a negative coupling). In the former case, linear perturbations are only mildly suppressed or almost unaffected, while in the latter they result significantly suppressed.
At nonlinear scales, however, the time variation of the extra fifth-force becomes the dominant effect giving rise to an excess of power and a significant increase of the concentration of CDM halos for models with a negative coupling, as these feature a positive derivative of the effective gravitational constant at low $z$, while for positive couplings (i.e. a decreasing effective gravitational constant) the opposite effect arises. For all cases, the HMF is found to be suppressed as compared to $\Lambda $CDM. 

Once taking into account the fact that \citeauthor{Li_Mota_Barrow_2011a} adopted the same initial conditions for all their different DE simulations, thereby implicitly imposing a common normalization of the linear perturbations amplitude of all the models at some intermediate redshift $z_{i}$ between last scattering and the present time, their results appear to be in general qualitative agreement with the earlier outcomes of \cite{DeBoni_etal_2011}.

As a follow-up to their early work, \cite{DeBoni_etal_2012} extended the analysis of their simulations to a detailed investigation of the concentration-mass relation for clusters in DE models including Extended Quintessence scenarios, finding again that the sign of the time derivative of the effective gravitational constant drives the shift in the normalization of the concentration-mass relation, consistently with the outcomes of \cite{Li_Mota_Barrow_2011a}. This effect is somewhat similar to the one detected for Coupled DE models with a variable coupling \citep[][see above and Fig.~\ref{fig5}]{Baldi_2011a} where the time variation of the effective gravitational constant alters the virial equilibrium of collapsed objects inducing a contraction or an expansion of the halos and consequently an increase or a decrease of their concentration parameter as compared to $\Lambda $CDM.\\

The most general case of a universal interaction between a cosmic inhomogeneous scalar and matter fields in the Universe corresponds to cosmological models where nonlinearities in the function $F(\delta \phi )$ appearing in Eq.~\ref{modified_poisson} induce large spatial fluctuations in the scalar field, capable to provide an efficient screening of the extra fifth-force at small scales even for effective coupling values of order unity, i.e. for a strength of the fifth-force comparable to gravity. If significant nonlinearities in the function $F$ are present, in fact, it is no longer possible to discard the term $F(\delta \phi )$ in Eq.~\ref{modified_poisson} and approximately relate the scalar field perturbations $\delta \phi $ to the matter perturbaitons $\delta \rho _{\rm M}$ through a standard linear Poisson equation. This is the case of modified gravity models as e.g. $f(R)$ theories, Symmetron fields, or higher-dimensional theories of gravity that were introduced in Section~\ref{sec:models} above.
In these classes of models, then, it is strictly necessary to solve the full Eq.~\ref{modified_poisson} in order to compute the actual fifth-force acting on massive particles at different positions, without resorting on any further approximation, as the behavior of the extra force will be different in different environments due to the explicit screening mechanisms defined by the details of the function $F(\delta \phi )$ and of the coupling function $\beta (\phi )$. This is an extremely challenging task in itself, which requires dedicated algorithms to be included and interfaced with standard N-body solvers, and that sensibly increases the computational cost of large N-body runs for these scenarios.

The field of cosmological simulations of modified gravity models, and in general of scalar field cosmologies with explicit screening mechanisms, is still rather young, although in recent years the efforts to develop competitive and versatile N-body codes for this class of scenarios have been remarkable. A proper review of such field is then probably still premature, since it is only very recently that independent simulation codes have started to produce broadly consistent results for some specific realizations of modified gravity theories, and a proper comparison of different algorithms for cross-checking and mutual validation has not yet been performed. 
Nevertheless, the amount of work invested by several different research groups in developing and testing such implementations over the last few years is definitely worth a mention, along with the clear prediction that in a relatively short timescale a large amount of robust and significant results in this field will be achieved through a systematic program of numerical investigations. I will therefore provide here only a very brief and general overview of this field, and leave to a future time a more thorough discussion.\\

\begin{figure*}
\includegraphics[height=2.4in]{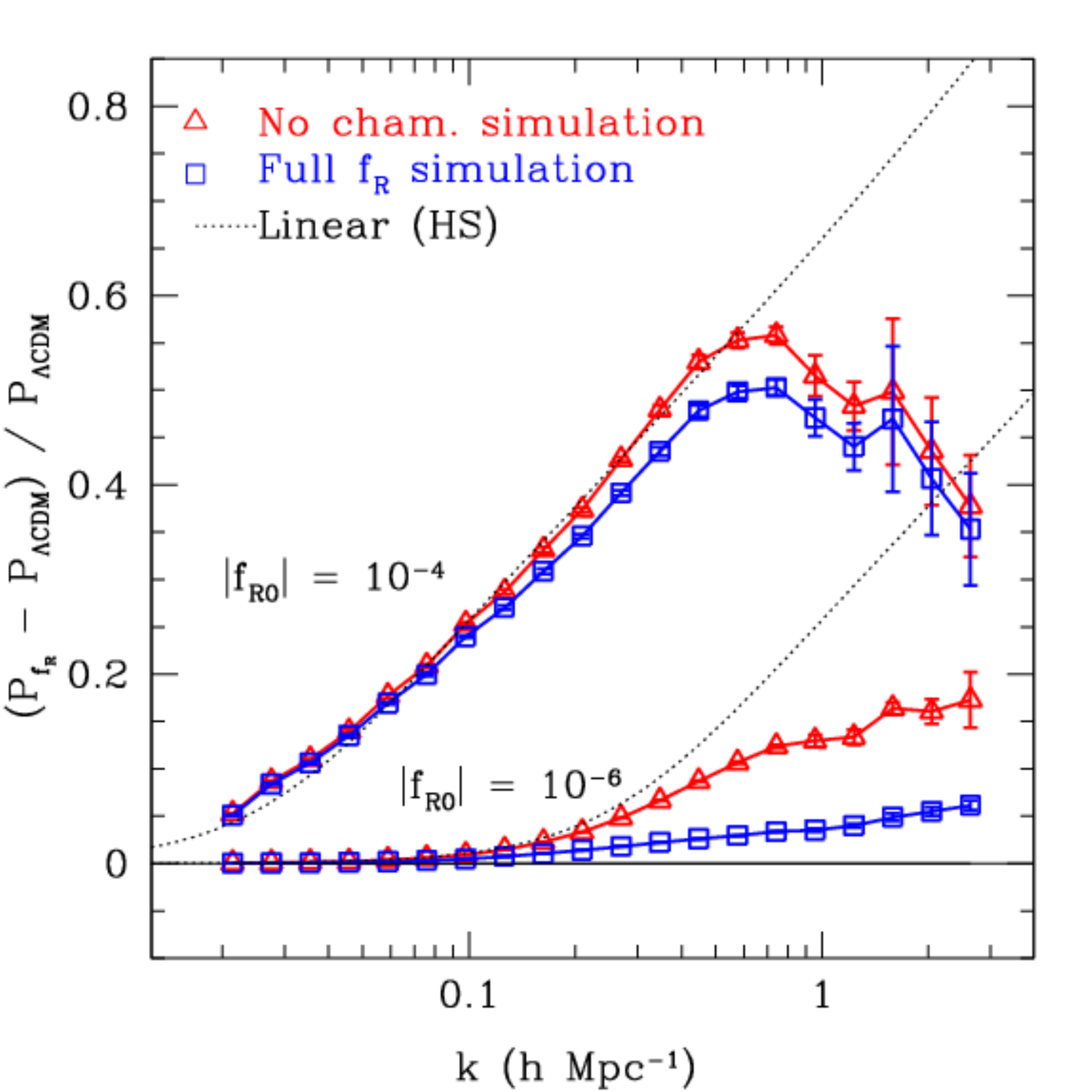}
\includegraphics[height=2.6in]{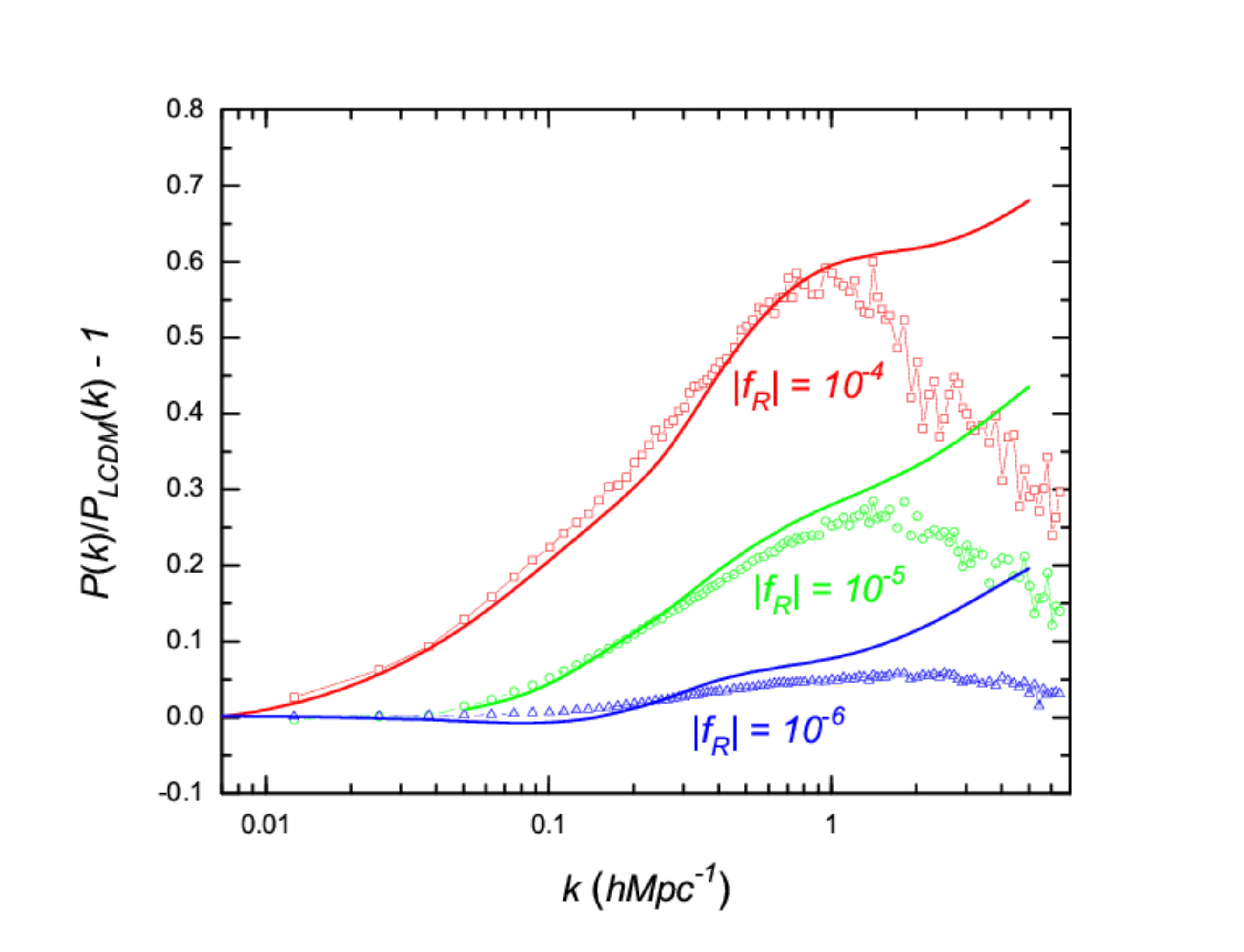}
\caption{The effect of a series of $f(R)$ modified gravity theories on the nonlinear matter power spectrum as compared to the standard $\Lambda $CDM cosmology within General Relativity. The plots are extracted from the first simulations of $f(R)$ models by \cite{Oyaizu_etal_2008} ({\em Left}) and from the more recent results of \cite{Ecosmog} ({\em Right}). The enhancement in the nonlinear matter power spectrum due to the additional fifth-force related to the modified gravity theory is significantly suppressed at small scales by the Chameleon screening mechanism, in a way that appears consistent between these two different implementations of $f(R)$ gravity models in N-body simulations.}
\label{fig8}
\end{figure*}

The first attempt to run N-body simulations for modified gravity in the form of $f(R)$ theories was made by \cite{Oyaizu_2008,Oyaizu_etal_2008,Schmidt_etal_2009}, who made use of an iterative relaxation scheme on a fixed cartesian grid within a mesh-based N-body code to solve Eq.~\ref{modified_poisson} and self-consistently compute the motion of particles in the presence of a modified gravity law given by the superposition of the standard gravitational interaction and a screened fifth-force. Their results showed for the first time how the screening mechanism strongly suppresses the effects of the fifth-force at small scales, and in particular in the inner parts of collapsed halos where the suppression is most effective (see Fig.~\ref{fig8}, left panel). A similar numerical approach was followed soon after by \cite{Li_Zhao_2009,Zhao_etal_2010,Zhao_Li_Koyama_2011a} who implemented a Newton-Gauss-Seidl iterative solver for the scalar field nonlinear Poisson equation on the adaptive grid of the AMR code {\small MLAPM}, allowing for a significant improvement of the code resolution as compared to the earlier implementation of \citeauthor{Oyaizu_2008} in regions where the screening mechanism is most efficient. The relatively small simulations that could be carried out with such AMR implementation of modified gravity gave results in good agreement with previous findings, and triggered a significant number of post-processing analysis aimed at investigating possible characteristic features of modified gravity models in various observables. {These include e.g.} the statistics of CDM halos and voids \citep[as in][]{Zhao_etal_2010,Zhao_Li_Koyama_2011a,Winther_Mota_Li_2011,Li_Zhao_Koyama_2011}, the geometrical and dynamical properties of virialized halos \citep[][]{Lombriser_etal_2012,Lee_etal_2012, Llinares_Mota_2012} or the apparent variation of the fine-structure constant \citep[as in][]{Li_Mota_Barrow_2011b} as well as a number of applications to different types of screening mechanisms such as the Dilaton \citep[][]{Brax_etal_2011} and the Symmetron \citep[][]{Davis_etal_2012}. The same implementation has been recently ported by \cite{Ecosmog} into the hydrodynamical AMR N-body code {\small RAMSES} and has been named {\small ECOSMOG} standing for Efficient COde for Simulating MOdified Gravity. Such code overcomes several shortcomings of the previous {\small MLAPM} implementation concerning the parallelization strategy and the multi-grid refinement for solving the nonlinear Poisson equation, which makes the {\small ECOSMOG} code more suitable for large simulations with high mass resolution. The first series of simulations performed with such code have been focused on
$f(R)$ models characterized by a {\em Chameleon} screening mechanism as well as on different types of screening as e.g. {\em Symmetron-} and {\em Dilaton-}type modified gravity theories \citep[see e.g.][]{Brax_etal_2012}, and are now starting to be post-processed and analyzed with a particular focus on the effects of modified gravity theories on the nonlinear matter and velocity power spectra \citep[][]{Li_etal_2012} and their connection with direct observables such as the detailed pattern of redshift-space distortions in wide galaxy surveys \citep[][]{Jennings_etal_2012}, finding consistent results with previous works (see Fig.~\ref{fig8}, right panel).

The number of independent implementations of modified gravity theories into cosmological N-body codes and the range of accurate predictions for direct observable quantities that are being produced with such last generation of N-body solvers seems encouraging in view of the needs of large upcoming surveys aimed at investigating the nature of the dark sector and of the gravitational interaction at cosmological scales.

\section{Simulating large-void cosmological models}
\label{sec:LTB}

As a last class of models of the cosmic acceleration beyond $\Lambda $ that have been
investigated by means of N-body simulations it is worth to mention the case of large-void
inhomogeneous cosmologies introduced in Section~\ref{sec:models}. The only cosmological simulations
of such models so far have been performed by \cite{Alonso_etal_2010} using the TreePM code {\small GADGET}. Such simulations did not require any specific modification of the public version of the {\small GADGET-2} code but rather a modification of the initial conditions generator to include the gravitational potential of a large void in the computation of the initial particles' displacements. This has been done by suitably modifying the code {\small 2LPT} based on 2nd order Lagrangian perturbation theory to include in the initial conditions a series of large voids with different values of the structural parameters $\Omega _{\rm in}$ and $\Delta r/r_{0}$, that have then been evolved using the standard newtonian N-body approach. One of the most relevant results of such investigation has been to demonstrate that standard N-body codes are suitable to correctly follow the nonlinear evolution of the density void (i.e. to correctly predict its evolution up to a density contrast $\delta \approx 1$ at the present time) without requiring any general relativistic modifications. Also, the same work showed that the linear density contrast in such inhomogeneous cosmologies is always very close to the standard $\Lambda $CDM prediction. Such results support the employment 
of large cosmological newtonian N-body simulations also to investigate scenarios for which the basic assumption of large-scale homogeneity encoded by the Copernican principle does no longer hold.

\section{Conclusions}
\label{sec:concl}

The last decades of investigation in the fields of Cosmology, Astrophysics, and Particle Physics, have provided us with a clear and undeniable quantification of our ignorance: about 96\% of the energy density in the Universe is made of particles and fields that keep eluding all our efforts of detection and identification. In such context, Dark Energy and Dark Matter are therefore simply labels that allow us to organize and classify the limited observational knowledge that we have been growing through the years about such ``dark side" of the Universe. Although the simplest and most widely accepted cosmological model -- that associates Dark Energy to a cosmological constant and Dark Matter to a family of Weakly Interacting Massive Particles beyond the standard model of particle physics -- is presently consistent with all our available observational data, its theoretical roots are difficult to accommodate in the context of General Relativity and Quantum Field Theory, and alternative scenarios keep being proposed almost on a daily basis since more than a decade. 

We are therefore accumulating an increasing number of alternative cosmological models that aim at providing a solution to the mystery of the fundamental nature of the dark Universe, which are often barely distinguishable from each other in their predictions concerning the background evolution of the Universe or the growth of linear density perturbations.
Trying to exploit also the nonlinear regime of structure formation as a possible way to discriminate among different cosmological scenarios and as a source of observational information about the nature of the dark Universe is therefore becoming a necessary further step in the connection between theory and observations in cosmology. Such a step however requires to make use of large numerical simulations as the nonlinearities involved in the problem prevent to drive any reliable conclusion based only on analytical tools.
This need has driven the wide range of efforts that have been made in the last years to develop, test, and ultimately apply new and highly sophisticated algorithms within N-body codes to self-consistently simulate the evolution of cosmic structures in the context of different and competing cosmological scenarios.\\

In this Review, I tried to provide a broad overview on the results of such new and rapidly developing research field, mainly focusing on cosmological N-body simulations
of Dark Energy models alternative to the standard $\Lambda $CDM scenario. After briefly reviewing the history of the role played by N-body simulations in establishing the
present standard cosmological model, I provided a broad (and necessarily incomplete) overview of the different Dark Energy scenarios that are presently being considered as 
possible competitors to the standard model. In doing so, I classified Dark Energy models in two different categories defined by the clustering properties of the Dark Energy field
(whatever this field might be) at sub-horizon scales, deliberately avoiding any attempt to make a fundamental distinction between Dark Energy and Modified Gravity scenarios. 
In fact, no fundamental distinction between a Dark Energy component in the stress-energy tensor of the Universe and a modification of the laws of gravity is possible when one allows for density perturbations and direct interactions of the Dark Energy field. 
The same classification, however, results also particularly useful to discuss the specific modifications that have to be implemented in cosmological N-body algorithms to account for the characteristic features of different Dark Energy models. In fact, depending on the clustering properties of the Dark Energy field, structure formation can be affected either
only through a modified background expansion history, or also by additional forces related to the local Dark Energy density perturbations. The numerical investigation of these two distinct possibilities through N-body simulations has been the main focus of the present Review. A further possibility, which does not belong
to any of these two main categories, is given by models that relate the observed accelerated expansion of the Universe to a local deviation form homogeneity. Such option requires a different kind of numerical implementation, and has been discussed separately. \\

In the second part of this work, I attempted a general overview of the main investigations that have been performed in the last decade using N-body simulations of non-standard Dark Energy models, following the general classification summarized above, and focusing the discussion only on the main outcomes of the various studies rather then on their technical details. Due to the complexity of the field, and to the wide range of different cosmological models, N-body codes, and normalization choices assumed in different studies, the description of most of the mentioned works has necessarily been incomplete and oversimplified, but I tried to highlight the most relevant results obtained by different research groups and to which extent such results have been subsequently confirmed or disproved by other independent investigations. In any case, I tried to provide an extensive list of references to address interested readers to the relevant literature.

In this broad overview, I mainly focused on homogeneous Dark Energy models and on various types of interacting Dark Energy cosmologies, that are the classes of models for which a wide number of independent cosmological simulations have been carried out so far, providing consistent results which can then be considered sufficiently robustly established.
In the last part of the Review, however, I provided also a brief and necessarily incomplete summary of the main efforts that have been put in place in the last years to develop N-body codes for various types of Modified Gravity models, that according to the above-mentioned classification scheme correspond to inhomogeneous Dark Energy models with a universal screened interaction to massive particles. The field of N-body simulations of Modified Gravity models is presently very active and has shown an impressive progress in the last couple of years,
but the very recent development and application of most of the presently available codes together with the lack of a direct comparison of different implementations make a proper review of this specific field probably still premature.\\

The number of different and independent efforts aimed at developing suitable numerical tools to push the comparison between theoretical models of the dark side of the Universe
and direct observations deep into the nonlinear regime of structure formation has been continuously growing in the last decade. Despite the difficulties related to the intrinsic nonlinear nature of the processes under investigation, the field of N-body simulations of Dark Energy and Modified Gravity models seems to be rapidly developing and promises to 
provide highly constraining  predictions for a wide range of presently viable and competing cosmological scenarios. The additional complications, which have not been discussed in the present work, related to possible degeneracies with other 
physical processes in place at the same nonlinear scales at which present N-body simulations of non-standard cosmologies are making their most valuable predictions, will necessarily have to be taken in full consideration in the future. {Also,} a synergy between a proper implementation of Dark Energy models and a better understanding of baryonic physics will be required to obtain the level of accuracy which is demanded by the next generation of observational surveys. \\

Although an impressive range of results have been obtained in the last decade from N-body simulations of non-standard cosmological scenarios, we are only at the beginning of a long and challenging path for numerical cosmology, and the present work is probably only the first of a long series of Reviews in this new and exciting field of research.

\section*{Acknowledgements}
I would like to thank the two anonymous Referees for valuable comments on the manuscript.
This work has been supported by the DFG Cluster of Excellence ``Origin and Structure of the Universe'' and by
the TRR33 Transregio Collaborative Research Network on the ``Dark%
Universe''.

\footnotesize
\bibliographystyle{elsarticle-harv}
\bibliography{baldi_bibliography.bib}







\end{document}